\documentclass[pre,twocolumn,superscriptaddress]{revtex4}

\usepackage{graphicx}
\usepackage{amssymb,amsfonts,amsmath}
\usepackage{color}
\usepackage{mathrsfs}
\usepackage[T1]{fontenc}

\begin{document}

\title{Leveraging large-deviation statistics to decipher the stochastic
properties of measured trajectories}

\author{Samudrajit Thapa}
\affiliation{Institute for Physics \& Astronomy, University of Potsdam, 14476
Potsdam-Golm, Germany}
\author{Agnieszka Wy{\l}oma{\'n}ska} 
\affiliation{Faculty of Pure and Applied Mathematics, Hugo Steinhaus Center,
Wroc{\l}aw University of Science and  Technology, Wroc{\l}aw, Poland}
\author{Grzegorz Sikora}
\affiliation{Faculty of Pure and Applied Mathematics, Hugo Steinhaus Center,
Wroc{\l}aw University of Science and  Technology, Wroc{\l}aw, Poland}
\author{Caroline E. Wagner}
\affiliation{Princeton Environmental Institute, Princeton University, Princeton NJ
08544, USA}
\affiliation{Department of Ecology and Evolutionary Biology,
Princeton University, Princeton NJ 08544, USA}
\author{Diego Krapf}
\affiliation{Department of Electrical and Computer Engineering, Colorado State
University, Fort Collins, Colorado 80523, USA}
\affiliation{School of Biomedical Engineering, Colorado State University, Fort Collins,
Colorado 80523, USA}
\author{Holger Kantz}
\affiliation{Max Planck Institute for the Physics of Complex Systems, Dresden, Germany}
\author{Aleksei V. Chechkin}
\affiliation{Institute for Physics \& Astronomy, University of Potsdam, 14476
Potsdam-Golm, Germany}
\affiliation{Akhiezer Institute for Theoretical Physics National Science Center
"Kharkov Institute of Physics and Technology", Akademicheskaya st.1, Kharkov 61108,
Ukraine}
\author{Ralf Metzler}
\affiliation{Institute for Physics \& Astronomy, University of Potsdam, 14476
Potsdam-Golm, Germany}
\email{rmetzler@uni-potsdam.de}

\begin{abstract}
Extensive time-series encoding the position of particles such as viruses,
vesicles, or individual proteins are routinely garnered in single-particle
tracking experiments or supercomputing studies. They contain vital clues
on how viruses spread or drugs may be delivered in biological cells. Similar
time-series are being recorded of stock values in financial markets and of
climate data. Such time-series are most typically evaluated in terms of
time-average mean-squared displacements, which remain random variables for
finite measurement times. Their statistical properties are different for
different physical stochastic processes, thus allowing us to extract
valuable information on the stochastic process itself. To exploit the
full potential of the statistical information encoded in measured time-series
we here propose an easy-to-implement and computationally inexpensive new
methodology, based on deviations of the time-averaged mean-squared displacement
from its ensemble average counterpart. Specifically, we use the upper bound of
these deviations for Brownian motion to check the applicability of this approach
to simulated and real data sets. By comparing the probability of deviations for
different data sets, we demonstrate how the theoretical bound for Brownian motion
reveals additional information about observed stochastic processes. We apply the
large-deviation method to data sets of tracer beads tracked in aqueous solution,
tracer beads measured in mucin hydrogels, and of geographic surface temperature
anomalies. Our analysis shows how the large-deviation properties can be
efficiently used as a simple yet effective routine test to reject the Brownian
motion hypothesis and unveil crucial information on statistical properties such as
ergodicity breaking and short-time correlations.
\end{abstract}

\date{\today}

\maketitle

\section{Introduction}

Brownian Motion (BM) is characterized by the linear scaling with time
of the mean squared displacement (MSD), $\langle\mathbf{r}^2(t)\rangle=2dDt$
in $d$ dimensions, where $D$ is the diffusion coefficient and angular brackets
denote the ensemble average over a large number of particles. In many biological
and soft-matter systems, this linear scaling has been reported to be violated
\cite{hoff13,lenerev,krapf11,tabei}. Instead, \emph{anomalous\/} diffusion with
the power-law scaling $\langle\mathbf{r}^2(t)\rangle\simeq t^\alpha$ of the MSD
is observed. The anomalous diffusion exponent $\alpha$ characterizes
\emph{subdiffusion\/} when $0<\alpha<1$ and \emph{superdiffusion\/} when $\alpha
>1$ \cite{metz12,igorsoft,metz14}.

Passive and actively-driven diffusive motion are key to the spreading of viruses,
vesicles, or proteins in living biological cells \cite{seisenhuber,christine,
gratton}. Pinpointing the precise details of their dynamics will ultimately pave
the way for improved strategies in drug delivery, or lead to better understanding
of molecular signaling used in gene silencing techniques. Similarly, improved
analyses of the stochastic dynamics of financial or climate time series will
allow us to find better comprehension of economic markets or climate impact.

The most-used observable in the analysis of time-series $\mathbf{r}(t)$ garnered
for the position of viruses or vesicles by modern single-particle tracking setups
in biological cells or for the key quantities in financial or climate dynamics,
such as price or temperature, is the time-averaged MSD (TAMSD)
\cite{metz12,metz14}
\begin{equation}
\overline{\delta^2(\Delta)}=\frac{1}{T-\Delta}\int_0^{T-\Delta}\Big[\mathbf{r}(
t+\Delta)-\mathbf{r}(t)\Big]^2dt,
\label{eq-normal-time-ave}
\end{equation}
expressed as function of the lag time $\Delta$. For BM at sufficiently long $T$,
the TAMSD \eqref{eq-normal-time-ave} converges to the MSD, formally $\lim_{T\to
\infty}\overline{\delta^2(\Delta)}=\langle\mathbf{r}^2(\Delta)\rangle=2dD\Delta$,
reflecting the \emph{ergodicity\/} of this process in the Boltzmann-Khinchin
sense \cite{stas}. Anomalous diffusion processes may be MSD-ergodic,
with a TAMSD of the form $\overline{\delta^2(\Delta)}\simeq\Delta^{\beta}$
with $\beta=\alpha$, e.g., fractional Brownian motion (FBM), or they may be
"weakly non-ergodic", e.g., $\beta=1$ for continuous time random walks (CTRWs)
with scale-free waiting times \cite{metz12,metz14,stas}.

Due to the random nature of the process, the TAMSD is inherently irreproducible
from one trajectory to another, even for BM. The emerging amplitude spread is
quantified in terms of the dimensionless variable $\xi=\overline{\delta^2(\Delta)}
/\left<\overline{\delta^2(\Delta)}\right>$, where $\left<\overline{\delta^2(\Delta)}
\right>$ is the average of the TAMSD over many trajectories \cite{stas,metz14}. The
variance of $\xi$ is the ergodicity breaking parameter $\mathrm{EB}(\Delta)=\left<
\xi^2\right>-1$. Together with the full distribution $\phi(\xi)$, EB provides
valuable information on the underlying stochastic process \cite{metz14}. For BM,
in the limit of large $T$, each realization leads to the same result, $\phi(\xi)
=\delta(\xi-1)$ and $\mathrm{EB}=0$. For scale-free CTRWs, even in the limit $T
\to\infty$ EB retains a finite value and the TAMSD remains a random variable,
albeit with a known distribution $\phi(\xi)$ \cite{metz12,metz14,stas}.

The MSD and TAMSD or, alternatively, the power spectrum and its single trajectory
analog \cite{dieg18,dieg19}, are insufficient to fully characterize a measured
stochastic process. A TAMSD of the form $\overline{\delta^2(\Delta)}\simeq\Delta$,
e.g., may represent BM or weakly non-ergodic anomalous diffusion. Similarly, the
linearity of the MSD, $\langle\mathbf{r}^2(t)\rangle\simeq t$ is the same for BM
and for random-diffusivity models with non-Gaussian distribution (see below). For
the identification of a random process from data, additional observables need to
be considered which may then be used to build a decision tree \cite{igor}. Recent
work targeted at objective ranking of the most likely process behind the data is
based on Bayesian-maximum likelihood approaches or on machine learning applications
\cite{samu18,gorka19,bayes,bayes1}.
The disadvantage of these methods is that they are often
technically involved and thus require particular skills, plus computationally
expensive. Here we provide an easy-to-implement reliable method based on
large-deviation properties encoded in the TAMSD.
As we will see, this method is very delicate and able to identify
important properties of the physical process
underlying the measured data. Moreover, it detects correlations in the data and
has significantly sharper bounds than the well known Chebyshev inequality
\cite{cheby0,savage} widely used in different applications \cite{cheby,cheby1,cheby2}.
In the following we report analytical results for the large-deviation statistic of the
TAMSD and demonstrate the efficacy of this approach for various data sets ranging
from microscopic tracer motion to climate statistics.

\section{Large deviations of the TAMSD}

Large-deviation theory is concerned with the asymptotic behavior of large
fluctuations of random variables \cite{Cra38,Don1,Don4,feng}.
It finds applications in a wide range of fields such as information theory
\cite{dembo}, risk management \cite{novak}, or the development of sampling
algorithms for rare events \cite{bouchet18}. In thermodynamics and statistical
mechanics, large-deviation theory finds prominent applications as described in
\cite{touc09}. More recently large-deviations for a variety of random variables
have been analyzed for different stochastic processes \cite{sp1,spa2,spa3,sp5,spa1,eli}.
In fact large-deviation theory is closely related to extreme value statistics
\cite{extreme,extreme1,extreme2} (see also Appendix \ref{connect}).

An intuitive definition of the large-deviation principle can be given as follows.
Let $A_N$ be a random variable indexed by the integer $N$, and let $P(A_N\in B$)
be the probability that $A_N$ takes a value from the set $B$. We say that $A_N$
satisfies a large-deviation principle with rate function $I_B$ if $P(A_N\in B)
\approx e^{-NI_B}$ \cite{touc09}. The exact definition operates with supremum
and infimum of the above probability and the rate function \cite{feng}.
However, sometimes it is difficult or even impossible to find explicit formulas
for the rate function or the large-deviation principle. Still, in such cases
one may be able to find an upper bound for the probability $P(A_N\in B)$,
i.e., the function $I_B(N)$ which satisfies $P(A_N\in B)\leq e^{-I_B(N)}$. This
is exactly the case we consider here.

When the TAMSD is a random variable and we have expressions for $I_B$ corresponding
to specific processes, we arrive at upper bounds on the probability, $P((\xi-1)>
\varepsilon)$ that a given realization of the TAMSD deviates from the expected mean by
a preset amount $\varepsilon$: $P\left((\xi-1)>\varepsilon\right)\leq e^{-I(\varepsilon,
\Delta,N)}$. Here, $I$ is a function of the deviation $\varepsilon$, the lag time
$\Delta$, and the number $N$ of points in the trajectory.

\subsection*{Theoretical bounds on the deviations of TAMSD}

BM is characterized by the overdamped Langevin equation $dX(t)/dt=
\sqrt{2D}\eta(t)$, driven by white Gaussian noise $\eta(t)$ with
zero mean and autocorrelation function $\langle\eta(t_1)\eta(t_2)\rangle=\delta(
t_1-t_2)$. In the following we consider discretized trajectories of BM, $\mathbb{
X}=(X(1),X(2),\ldots,X(N))$. For BM the following statements can be shown to
hold.

\subsubsection{Chebyshev's inequality}

Before we come to large-deviations, we recall the (one-sided) Chebyshev inequality
for any random variable $X$ with mean $\mu$ and finite variance.For BM, Chebyshev's
inequality for the TAMSD reads (see Appendix \ref{sec-chebyshev-derivation} for
details)
\begin{eqnarray}
\label{eq-cheb-final}
P\left((\xi-1)\ge\varepsilon\right)\leq4\Delta/(4\Delta+3N\varepsilon^2).
\end{eqnarray}
While this inequality is useful for a first analysis and will serve as a
reference below, we will show that the large-deviation result
presented here has significantly sharper bounds.

\subsubsection{Large deviations of TAMSD for BM}

From large-deviation theory for BM, the following result can be derived
\cite{gadj18}
\begin{equation}
\label{eq-dev-tamsd-bm}
P\left((\xi-1)>\varepsilon\right)\leq\exp\left(-a\mathcal{H}\left(b\right)\right),
\end{equation}
where $a=[4(N-\Delta)D^2\Delta(\Delta+1)(2\Delta+1)]/[3\bar{\lambda}(\Delta)^2]$
and $b=[3\bar{\lambda}(\Delta)\varepsilon]/[2D(\Delta+1)(2\Delta+1)]$. Moreover,
$\mathcal{H}(u)=1+u-\sqrt{1+2u}$ and $\bar{\lambda}(\Delta)=2\max\{\lambda_j(
\Delta)\}$, where $\lambda_j(\Delta)$ ($j=1,2,\ldots,N-\Delta$) are the
eigenvalues of the $(N-\Delta)\times(N-\Delta)$ positive-definite covariance
matrix ${\Sigma}(\Delta)$ for the increment vector $\mathbb{Y}=(X(1+\Delta)-
X(1),X(2+\Delta)-X(2),\ldots,X(N)-X(N-\Delta))$. Note that although the diffusion
coefficient $D$ explicitly appears in \eqref{eq-dev-tamsd-bm} it cancels out both
in the function $\mathcal{H}(\cdot)$ and its prefactor, as $\bar{\lambda}$
contributes the factor $D$. It is noteworthy that $I$ is independent of the
diffusion coefficient $D$. This can be understood
intuitively, as different values of $D$ in the log-log plot of the TAMSD merely
shift the amplitude but leave the amplitude spread unchanged \cite{metz14,stas}.

\begin{figure*}
\includegraphics[width=18cm]{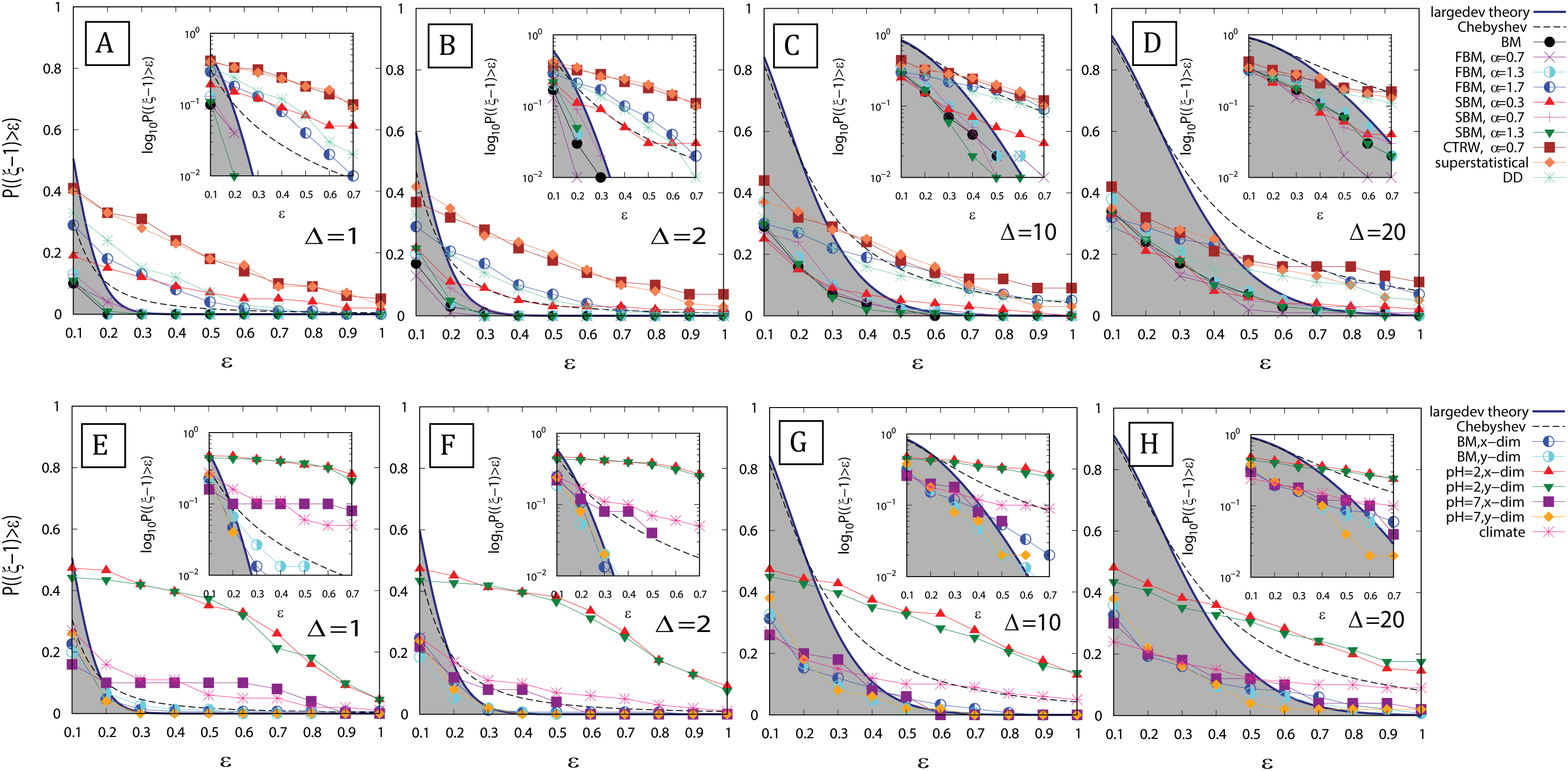}
\caption{Variation of the estimates of $P\left((\xi-1)>
\varepsilon\right)$ as function of the deviation $\varepsilon$. (A)-(D) show
results for simulated processes with $M=100$ trajectories of $N=300$ points each,
and for lag times $\Delta=
1$, 2, 10, and 20, respectively. The insets show the results on semi-log scale.
The statistical uncertainty is of the order of $0.01$. The parameters of the
simulated stochastic processes are: $D=0.5$ for BM, $D_H=0.5$ for FBM, $D_0=0.5$
for SBM, $\tau_0=1$ for CTRW, $D_0=10$ for the superstatistical process, $\tau_c
=10$ and $D_{\star}=0.2$ for DD. (E)-(H) show results for different
experimental datasets for lag times $\Delta=1$, 2, 10, and 20, respectively. The
insets again show the results on log-lin scale. The statistical uncertainty is of
the order of $0.01$. For more details Appendix \ref{simproc}.}
\label{fig1}
\end{figure*}

For the special choices $\Delta=1$ and $\Delta=2$ the eigenvalues of $\Sigma
(\Delta)$ can be calculated explicitly. This is relevant because for such low
values of $\Delta$, the conclusions drawn from the TAMSD analysis of sufficiently
long $T$ are statistically significant. For $\Delta=1$, the eigenvalues $\lambda_j
(\Delta=1)=2D$ and therefore $\bar{\lambda}=4D$. Using this in \eqref{eq-dev-tamsd-bm}
we get
\begin{equation}
\label{eq-dev-tamsd-bm-del1}
P\left((\xi-1)>\varepsilon\right)\leq\exp\{-(N-1)\mathcal{H}(\varepsilon)/2\}.
\end{equation}
For $\Delta=2$ the eigenvalues are given as (see Appendix \ref{eig}) $\lambda_j(
\Delta=2)=D[4+4\cos(j\pi/[N-1])]$. This expression can then be used to obtain
$\bar{\lambda}$ and thus $P\left((\xi-1)>\varepsilon\right)$ for $\Delta=2$. For
other values of $\Delta$, the eigenvalues are obtained numerically \cite{eigen}.

\section{Data sets for large-deviation analysis}

We here describe the data used in our analysis below. These contain both
BM and non-Brownian processes.

\subsection{Simulated data}

Simulated data serve as
benchmarks for the experimental data below. We simulate 100 trajectories each
for different processes (Fig.~\ref{fig1} A-D). This number of trajectories is
of the same order as in the experimental data sets. A larger set of 10,000
analyzed trajectories is presented in Fig.~\ref{figs1}. In addition to BM, we simulate
FBM, scaled Brownian motion (SBM), CTRW, superstatistical process, and
diffusing-diffusivity (DD) process, see Appendix \ref{simproc}
for their exact definition. FBM \cite{mand68} is governed by the
Langevin equation, driven by power-law correlated fractional Gaussian noise (FGN)
$\eta_{\mathrm{H}}(t)$ with Hurst index $H$ ($0<H<1$), related to the anomalous
diffusion exponent by $\alpha=2H$. SBM is characterized by the standard Langevin
equation but with time-dependent diffusivity $D(t)\propto t^{\alpha-1}$
\cite{metz14,lim}. CTRW is a renewal process with Gaussian jump lengths and
long-tailed distribution $\psi(\tau)\simeq\tau^{(-1-\alpha)}$ ($0<\alpha<1$) of
sojourn times between jumps \cite{scher75,montroll69}. For the simulated
superstatistical
process \cite{beck03,beck06} the diffusivity for each trajectory is drawn from a
Rayleigh distribution. Finally, the DD process is governed by the Langevin
equation with white Gaussian noise, but with a time-dependent, stochastic
diffusivity, evolving as the square of an Ornstein-Uhlenbeck process with
correlation time $\tau_c$ \cite{seno17}.

\subsection{Beads tracked in aqueous solution}

This data set (labeled "BM,
x-dim" and "BM, y-dim" for the two directions) consists of 150
two-dimensional trajectories from single particle tracking of 1.2 $\mu$m-sized
polystyrene beads in aqueous solution \cite{dieg18}. The time resolution of the
data is 0.01 sec.

\subsection{Beads tracked in mucin hydrogels}

These data are from micron-sized
tracer beads tracked in mucin hydrogels (MUC5AC with 1 wt\% mucin) at pH=2
(labeled "pH=2, x-dim"
and "pH=2, y-dim") and pH=7 (labeled "pH=7, x-dim" and "pH=7, y-dim")
\cite{rubi17}. The imaging was performed at a rate of 30.3 frames per second.
The pH=2 data set consists of 131 two-dimensional trajectories of 300 points
each while the pH=7 data set consists of 50 trajectories of 300 points each.

\subsection{Climate data}

We also use daily temperature records over a 100 year
period, after removing the annual cycle (these "anomalies" represent deviations
from the corresponding mean daily temperature) \cite{mess16}. This data set
consists of uninterrupted daily temperature recordings starting 1 January 1893
and are validated by the German Weather Service [Deutscher Wetterdienst (DWD),
2016]. The records were taken at the meteorological station at Potsdam
Telegraphenberg (52.3813 latitude, 13.0622 longitude, 81 m above sea level).

\begin{figure*}
\includegraphics[width=18cm]{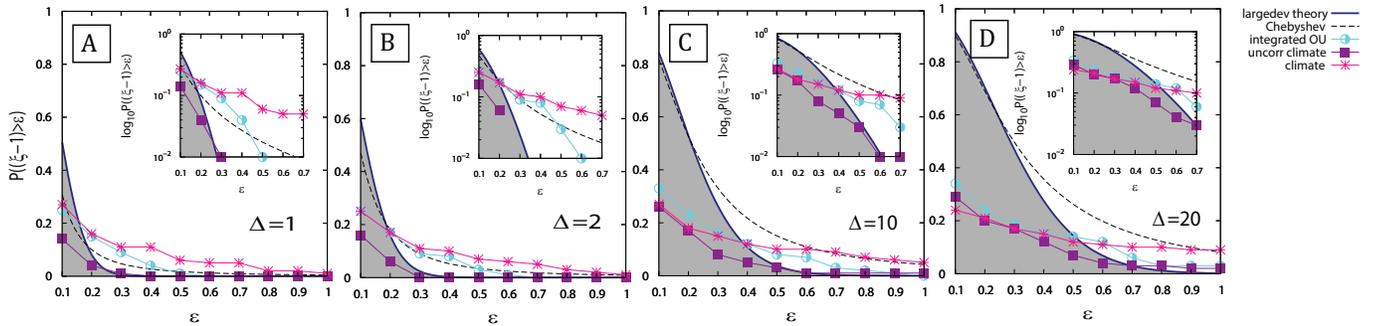}
\caption{Variation of the estimates of $P\left((\xi-1)>\varepsilon\right)$ as
function of $\varepsilon$ for the climate data, in comparison with the integrated
OU process with correlation time of 5 steps. (A)-(D), respectively, show the results
for $\Delta=1$, 2, 10, and 20. The insets show the results in semi-log scale. We
observe that random shuffling of the temperature anomalies, before taking the
cumulative sum to create the trajectories, removes the correlations in the data,
and $P((\xi-1)>\varepsilon)$ behaves very similarly to BM ("uncorr climate").}
\label{fig2}
\end{figure*}

\section{Results}

\subsection{Large deviations in simulated data sets}

Fig.~\ref{fig1} A-D shows the comparison of the
simulated data with the theoretical upper bounds \eqref{eq-cheb-final} and
\eqref{eq-dev-tamsd-bm} for BM, as function of the deviation $\varepsilon\in[0.1,1]$.
Here each of the $M=100$ simulated trajectories was of length $N=300$. We see that,
particularly at
short $\Delta$, the theoretical bound \eqref{eq-dev-tamsd-bm} from large-deviation
theory clearly distinguishes model classes and/or diffusive regimes. BM, subdiffusive
FBM, and superdiffusive SBM clearly lie below the bound \eqref{eq-dev-tamsd-bm}. In
contrast, superdiffusive FBM with a large $H$ exponent, subdiffusive SBM with a
small $\alpha$, CTRW, and random-diffusivity models
(superstatistical and DD) clearly exceed the bound \eqref{eq-dev-tamsd-bm}. Thus,
non-Gaussianity (as realized for CTRW and the random-diffusivity processes) is not
a unique criterion for the violation of the large-deviation bound. But according to
these results the large-deviation method, for a given value of the scaling exponent
$\alpha$, allows to distinguish FBM and SBM that both have a Gaussian PDF. At longer
$\Delta$, in general, the theoretical bound \eqref{eq-dev-tamsd-bm} increases and
thus $P\left((\xi-1)>\varepsilon\right)$ is below this bound for a larger range of
$\varepsilon$. The large deviation method is surprisingly robust with respect to
the number of analyzed trajectories, as can be seen from the marginal improvement
of the results based on 10,000 trajectories in Fig.~\ref{figs1}. Figs.~\ref{figs2}
and \ref{figs3} further analyze FBM and demonstrate the validity of the theoretical
bound (\ref{covmat}) derived for FBM. For further analysis of SBM see Fig.~\ref{figs4}.

Chebyshev's inequality (\ref{eq-cheb-final}) essentially provides the same
bound as the one from large-deviation theory for short $\Delta$. However,
at longer $\Delta$ it provides a much higher estimate than large-deviation
theory, and it is unable to distinguish subdiffusive SBM with $\alpha=0.3$
from BM, as both lie below this bound. Moreover, for long $\Delta$ the probability
$P\left((\xi-1)>\varepsilon\right)$ for all simulated processes lie either
below or very close to the bound of Chebyshev's inequality, rendering it
ineffective in discerning different processes.
Chebyshev's inequality \eqref{eq-cheb-final} lies above the large
deviation bound \eqref{eq-dev-tamsd-bm}, except for the cases $\Delta=1$ and 2
with small $\varepsilon$, when it is slightly below but still quite close to the
bound set by \eqref{eq-dev-tamsd-bm}.

\subsection{Large deviations in experimental data sets}

\subsubsection{Beads tracked in aqueous solution}

Polysterene beads tracked
in aqueous solution were analyzed in \cite{dieg18} using single-trajectory
power spectral analysis, concluding that the data are consistent with BM.
From Fig.~\ref{fig1} E-H it can be seen that the estimated probability
$P\left((\xi-1)>\varepsilon\right)$ somewhat exceeds the theoretical bound
\eqref{eq-dev-tamsd-bm} for BM. To understand this non-BM-like behavior
shown in the large-deviation analysis we closely examined the motion of
individual beads. Indeed, the displacement distributions of some beads
showed non-Gaussian behavior, that we could attribute to bead-bead
collisions as well as to imprecise localization of the bead center when
the recorded tracks suffered from non-localized brightness. We removed
the non-Gaussian trajectories using the JB test component-wise (see Appendix
\ref{sec-JB}). From the filtered data set ($M=129$ in $x$-direction and
$M=125$ in $y$-direction) we see that the large-deviation analysis within
the error bars is now consistent with BM (especially for $\Delta=1$, see
Fig.~\ref{figs9}). The large-deviation analysis is thus more sensitive to non-BM-like
behavior than other methods \cite{dieg18}. We also note that the analysis
based on Chebyshev's inequality could not distinguish these features.

\subsubsection{Beads tracked in mucin hydrogels}

The data sets ($M=131$ at
pH=2 and $M=50$ at pH=7) consisting of beads tracked in mucin hydrogels
show different trends of $P\left((\xi-1)>\varepsilon\right)$ depending on the
pH values, as seen for $N=300$ in Fig.~\ref{fig1} E-H. Notably, for the beads
tracked at pH=2 $P\left((\xi-1)>\varepsilon\right)$ remains significantly
above the bound set by \eqref{eq-dev-tamsd-bm}, particularly at short $\Delta$.
This implies that the spread of the TAMSD is inconsistent with BM and hence the
dynamics cannot be explained solely by BM. The data sets at pH=7 show
significantly different behavior. We observe a clear distinction in the trend
of $P\left((\xi-1)>\varepsilon\right)$ along the two directions of motion. Along
the direction (labeled "x-dim") $P\left((\xi-1)>\varepsilon\right)$ remains
slightly above the theoretical bound for BM from large-deviation theory for
most of the range of $\varepsilon$ at $\Delta=1$ and $\Delta=2$, while it remains
below the theoretical bound for the motion along the $y$-direction. As for the
beads in aqueous solution, Chebyshev's inequality provides a looser bound.

The mucin data sets were analyzed extensively in terms of Bayesian and
other standard data analysis methods in \cite{mucin19}. The MSD and
TAMSD exponents for the data at $\mathrm{pH}=2$ and 7 correspond to $\alpha=0.46$ and
0.36 and $\left<\beta\right>=1.09$ and 0.94, respectively. The
angular bracket for $\beta$ denotes that these exponents were determined
from the ensemble-averaged TAMSD. The discrepancy between the $\alpha$ and
$\beta$ values suggest ergodicity breaking and hence a contribution from a
model such as CTRW. For CTRW, the ensemble-averaged TAMSD scales with the total
measurement time $T$ as a power-law \cite{metz14}. However, as shown in \cite{mucin19},
the ensemble-averaged TAMSD for the data sets at $pH=7$ showed no dependence on $T$,
while the data sets at $pH=2$ showed a very weak dependence, ruling out
CTRW as a model of diffusion. Moreover in the Bayesian analysis carried
out in \cite{mucin19} BM, FBM and DD models were compared and relative
probabilities were assigned to each of them, based on the likelihood for each
trajectory to be consistent with a given process. It was observed that for both pH=2
and pH=7, and for most of the trajectories, both BM and FBM had high probabilities. On
comparing the estimated Hurst index $H$ for the FBM, it was seen that for pH=7,
$H\approx 0.5$ with a very small spread from trajectory to trajectory. In this
sense, the pH=7 data seemed to be very close to BM. This was also confirmed
independently by looking at $\beta$ extracted from the TAMSD. In contrast, the
estimated $H$ for the pH=2 data showed a large spread in the range $0.3\leq
H\leq0.7$. These observations are now clearly supported by the results for
$P\left((\xi-1)>\varepsilon\right)$, demonstrating that the data sets at $\mathrm{
pH}=7$ are close to BM while the data sets at $\mathrm{pH}=2$ cannot be explained
(solely) by BM. Thus, for this data set the large-deviation analysis again
demonstrates its effectiveness in unveiling the physical origin of the stochastic
time series.

\subsubsection{Climate data}

The climate data were successfully modeled
by an autoregressive fractionally integrated moving average model, more
specifically, ARFIMA(1,d,0) with $d\approx0.15$ \cite{mess16,kantz19}. ARFIMA(0,d,0)
corresponds to FGN with $H=d+0.5$. It was found that the data, in addition
to long-range correlations characteristic of FGN, exhibited short range
correlations due to which ARFIMA(1,d,0) fitted the data better than ARFIMA(0,
d,0). These short-range correlations could be explained by the average atmospheric
circulation period of 4-5 days \cite{mess16}. For our tests of deviations of the
TAMSD from the ensemble-averaged TAMSD, we construct FBM trajectories ($M=100$)
of length $N=300$ by taking a cumulative sum of FGN. If the temperature anomalies
could be described by ARFIMA(0,d,0), or, equivalently, by FGN, the cumulative sum
would be FBM and hence should show a similar trend of $P\left((\xi-1)>\varepsilon
\right)$, as seen for the simulated FBM processes in Fig.~\ref{fig1} A-D. That
means that it remains below the theoretical upper bounds \eqref{eq-cheb-final} and
\eqref{eq-dev-tamsd-bm} for FBM, as long as the scaling exponent does not become
too large. Alternatively, deviations of $P\left((\xi-1)>\varepsilon\right)$ from
the trend exhibited by simulated FBM, particularly at $\alpha=1.3$ corresponding
to $d=0.15$ reported in \cite{mess16}, would support the result in \cite{mess16}
that ARFIMA(0,d,0) does not completely explain the data of surface temperature
anomalies. This indeed turns out to be the case for $N=300$ in Fig.~\ref{fig1}
E-H where we observe that $P\left((\xi-1)>\varepsilon\right)$ remains \emph{above\/}
the theoretical upper bound for BM from large-deviation theory, especially
at short $\Delta$. Moreover, comparing with Fig.~\ref{fig1} A-D
we clearly observe that $P\left((\xi-1)>\varepsilon\right)$ remains well above
the theoretical upper bound \eqref{eq-dev-tamsd-bm} for BM for the climate
data at sufficiently large values of $\varepsilon$, while it always remains below
the bound for simulated FBM with $\alpha=1.3$ for all lag times. This corroborates
the finding in \cite{mess16} that ARFIMA(0,d,0) (or equivalently FBM for the
data constructed by taking the cumulative sum) cannot completely explain the
climate data. In comparison, Chebyshev's inequality \eqref{eq-cheb-final}
provides the same information for short lag times but fails to distinguish
the climate data from corresponding simulated FBM for long $\Delta$, as this bound
lies above the empirical probability $P\left((\xi-1)>\varepsilon\right)$ for both
corresponding simulated FBM and climate data. In order to check whether the short-term
correlations are indeed relevant, we create an artificially correlated process
in the form of an integrated Ornstein-Uhlenbeck (OU) process, the results of
which are shown in Fig.~\ref{fig2}. With a correlation length of five steps
the result of this OU process indeed leads to an $\varepsilon$ dependence of
$P\left((\xi-1)>\varepsilon\right)$ that is very similar to the climate data's.
Conversely, as
soon as we remove the correlations in the climate data by random reshuffling
of the temperature anomalies, the large-deviation behavior becomes BM-like.

\section{Discussion and Conclusion}

It is the purpose of time series analysis to detect the underlying physical
process encoded in a measured trajectory, and thus to unveil the mechanisms
governing the spreading of, e.g., viruses, vesicles, or signaling proteins
in living cells or tissues. Recently considerable work has been directed to
the characterization of stochastic trajectories using Bayesian analysis
\cite{lomh18,lomh17,samu18,bayes,bayes1} and machine learning
\cite{gorka19,eich19,
yael19}. Most of these methods are technically involved and expensive
computationally. Moreover, the associated algorithms often heavily rely on data
pre-processing \cite{gorka19}. To avoid overly expensive computations, it is
highly advantageous to first go through a decision tree, to narrow down the
possible families of physical stochastic mechanisms. For instance, one can
eliminate ergodic versus non-ergodic or Gaussian versus non-Gaussian processes,
etc. Here we analyze a new method based on large-deviation theory, concluding that
it is a highly efficient and easy-to-use tool for such a characterization. We
show how we can straightforwardly infer relevant information on the underlying
physical process based on the theoretical bounds of the deviations of the
TAMSD---routinely measured in single-particle-tracking experiments and
supercomputing studies and easy to construct for any time-series such as daily
temperature data---from the corresponding trajectory-average. Specifically, we
demonstrate that this tool is able to detect the short-time correlations
which effect non-FBM behavior in daily temperature anomalies, as well as the
crossover from BM-like behavior at pH=7 to non-ergodic, non-BM-like at
pH=2 for the mucin data, and the delicate sensitivity to non-Gaussian
trajectories for beads in aqueous solution.
We conclude from our analyses here that the large-deviation method would be
an excellent basis for a first efficient screening of measured trajectories,
before, if necessary, more refined methods are applied.

There are two seeming limitations to the large-deviation tool. First, it is
easy to formulate this tool for one-dimensional trajectories, while the
generalization to higher dimensions is not straightforward. However, as we
demonstrated it can be used component-wise and, remarkably, can be used to
probe the degree of isotropy of the data. In fact, from Fig.~\ref{fig1} E and
F we concluded that the tracer bead motion in mucin at pH=7 was
non-isotropic. In this sense, the one-dimensional definition of the large
deviation tool is in fact an advantage. Second, it is not trivial to derive
similar expressions as \eqref{eq-dev-tamsd-bm} for other stochastic processes.
Here, numerical evaluations can be used instead. Moreover, in this case we can
also use Chebyshev's inequality, with the caveat that it works best at short
lag times $\Delta$. Generally, the bound provided by the large-deviation
theory is considerably more stringent than Chebyshev's inequality, as
demonstrated here.

We demonstrated that superdiffusive FBM with large $H$ values is outside the 
large-deviation bound. Superdiffusive FBM applied in mathematical finance are
indeed in this range of $H$ values \cite{fin,fin1,fin2}, and our large-deviation
tool is therefore well suited for the analysis of such processes.
We also showed that the large-deviation tool is able to uncover subtle
correlations in the data, similarly to ARFIMA analyses applied mainly in
mathematical finance and time series analysis. This similarity between
the two methods strengthens the connections to physical models recently
worked out between random coefficient autoregressive models and
random-diffusivity models \cite{jakub}.

The large-deviation test investigated here is a highly useful
tool serving as an easy-to-implement and to-apply initial test in the decision
tree for the classification of the physical mechanisms underlying measured time
series from single particle trajectories.

\begin{acknowledgments}

S.T. acknowledges Deutsche Akademischer Austauschdienst (DAAD) for a PhD
Scholarship (program ID 57214224). C.E.W. is an Open Philanthropy Project
fellow of the Life Sciences Research Foundation. R.M. and A.C. acknowledge
financial support by the German Science Foundation (DFG, Grant ME 1535/7-1).
R.M. also acknowledges the Foundation for Polish Science (Fundacja na rzecz
Nauki Polskiej) for support within a Humboldt Polish Honorary Research
Scholarship.

\end{acknowledgments}

\begin{appendix}

\section{Description of the test algorithm}

We take $M$ discretized trajectories of length $N$ of a given process (simulated
or experimental). For the fixed time lag $\Delta$ we proceed as follows:
\begin{enumerate}
\item Calculate TAMSD for each trajectory according to the discrete equation,
$\overline{\delta^2(\Delta)}=\sum_{j=1}^{N-\Delta}\left(X(j+\Delta)-X(j)\right)^2$.
\item Calculate the ensemble-averaged TAMSD $\left<\overline{\delta^2(\Delta)}\right>$.
\item For each trajectory calculate $\xi=\frac{\overline{\delta^2(\Delta)}}{\left<
\overline{\delta^2(\Delta)}\right>}$.
\item Calculate the number of trajectories $M_{\epsilon}$ that satisfy the condition
$\xi-1>\epsilon$ for a given $\epsilon$. 
\item The empirical probability $P\left((\xi-1)>\epsilon\right)$ is calculated as
$M_{\epsilon}/M$.
\end{enumerate}
For a fixed value of $\epsilon$ and $\Delta$ we compare the empirical probability
$P\left((\xi-1)>\epsilon\right)$ with the theoretical bounds given by the
large-deviation theory and Chebyshev's inequality. In our analysis we consider
$\epsilon$ in the range $[0.1, 1]$, and $\Delta\ll N$, namely $\Delta=1,2,10$ and
$20$ points.

\section{Derivation of Chebyshev's inequality for TAMSD of BM}
\label{sec-chebyshev-derivation}

Using the Markov Inequality, one can also show that for any random variable
with mean $\mu$ and variance $\sigma^2$, and any positive number $k > 0$,
the following  Chebyshev inequality (one-sided) holds \cite{savage}
\begin{eqnarray}
\label{cheb}
P\left(X-\mu \geq k\right)\leq \frac{\sigma^2}{\sigma^2+k^2}.
\end{eqnarray}
Here we derive Chebyshev's inequality for the TAMSD statistic for BM. For the
TAMSD it takes the following form
\begin{eqnarray}
P\left((\xi-1)\geq k/\left<\overline{\delta^2(\Delta)}\right>\right)\leq\frac{
\sigma^2}{\sigma^2+k^2},\\
\nonumber
\end{eqnarray}
where $\sigma^2=\mathrm{Var}\left(\overline{\delta^2(\Delta)}\right)=4\left<
\overline{\delta^2(\Delta)}\right>^2\Delta/3N$, \cite{deng09}. Taking the notation 
$\varepsilon=k/\left<\overline{\delta^2(\Delta)}\right>$ one obtains the following
\begin{eqnarray}
\nonumber
P\left((\xi-1)\geq\varepsilon\right)&\leq&\frac{4\left<\overline{\delta^2(\Delta)}
\right>^2\Delta/3N}{4\left<\overline{\delta^2(\Delta)}\right>^2\Delta/3N+\epsilon^2
\left<\overline{\delta^2(\Delta)}\right>^2}\\
&=&\frac{4\Delta}{4\Delta+3N\epsilon^2}.
\label{cheb_final}
\end{eqnarray}

\section{Eigenvalues of the covariance matrix of increments for BM}
\label{eig}

\begin{widetext}
The $(N-\Delta)\times(N-\Delta)$ positive-definite covariance matrix ${\Sigma}
(\Delta)$ for the vector of increments $\mathbb{Y}=(X(1+\Delta)-X(1),X(2+\Delta)
-X(2),\ldots,X(N)-X(N-\Delta))$ takes the form
\begin{eqnarray}
\Sigma(\Delta) = 
\begin{bmatrix}
  \sigma_{\Delta}(0) & \ \ \ \sigma_{\Delta}(1)\ \ \ &\ \ \ \sigma_{\Delta}(2) \ \ \ & \ \ \ldots\ \  &\ \ \ \ldots \ \ \ &\sigma_{\Delta}(N-\Delta-1)  \\
  \sigma_{\Delta}(1) & \sigma_{\Delta}(0)  & \sigma_{\Delta}(1) &  \ddots   &  &  \vdots \\
  \sigma_{\Delta}(2)     & \sigma_{\Delta}(1) & \ddots  & \ddots & \ddots& \vdots \\ 
 \vdots &  \ddots & \ddots &   \ddots  & \sigma_{\Delta}(1) & \sigma_{\Delta}(2) \\
 \vdots &         & \ddots & \sigma_{\Delta}(1) & \sigma_{\Delta}(0)&  \sigma_{\Delta}(1) \\
\sigma_{\Delta}(N-\Delta-1)  & \ldots & \ldots & \sigma_{\Delta}(2)  & \sigma_{\Delta}(1) & \sigma_{\Delta}(0)
\end{bmatrix},
\end{eqnarray}
with its elements given by
\end{widetext}
\begin{eqnarray}\label{sigma_mala}
\sigma_\Delta(j)=\left\{ \begin{array}{ll}
2D(\Delta-j) & \textrm{$j\leq\Delta-1$}\\
0 & \textrm{$j>\Delta-1$}\\
\end{array} \right..
\end{eqnarray}
For the case of $\Delta=1$, the $(N-1)\times(N-1)$ covariance matrix ${\Sigma}(
\Delta=1)$ has elements given by
\begin{eqnarray}
\label{covmat}
\sigma_{\Delta=1}(j)=\left\{ \begin{array}{ll}
2D & \textrm{$j=0$}\\
0 & \textrm{$j>0$}\\
\end{array} \right..
\end{eqnarray}
Hence the matrix ${\Sigma}(\Delta=1)$ is a diagonal matrix with the
constant main diagonal $2D$ and all zero entries outside the main
diagonal. The characteristic polynomial of ${\Sigma}(\Delta=1)$ has the
form $$|{\Sigma}(\Delta=1)-\lambda I|=(\lambda-2D)^{N-1}$$ and roots
$\lambda_j(\Delta=1)=2D,$ which are the eigenvalues of that matrix.

For the case $\Delta=2$ the $(N-2)\times(N-2)$ covariance matrix ${\Sigma}(\Delta=2)$ has elements given by
\begin{eqnarray}
\sigma_{\Delta=2}(j)=\left\{ \begin{array}{ll}
2D(2-j) & \textrm{$j=0,1$}\\
0 & \textrm{$j>1$}\\
\end{array} \right..
\end{eqnarray}
Hence the matrix ${\Sigma}(\Delta=2)$ is a tridiagonal Toeplitz matrix. The formula forthe eigenvalues of such matrices is well known in the mathematical literature \cite{Bot05},
$$
 \lambda_j(\Delta=2)=D\left[ 4+4\cos\left(\frac{j\pi}{N-1}\right) \right].
$$

\section{Large deviations of TAMSD for FBM}

Taking Eq. (4.5) from \cite{gadj18} one can obtain the large deviation theory for FBM (see below for details of the stochastic process FBM). Namely, if we consider the  vector of increments $\mathbb{Y}=(X(1+\Delta)-X(1),X(2+\Delta)-X(2),\ldots,X(N)-X(N-\Delta))$ of FBM with Hurst exponent $H$ and generalized diffusion coefficient $D_H$ then we have
\begin{equation}\label{eq-dev-tamsd-fbm}
    P\left((\xi-1)>\epsilon\right) \leq \exp\left(- a \mathcal{H}\left( b \right) \right),	  
\end{equation}
where $a=\frac{-2(N-\Delta)D_{H}^2 S(\Delta,H,N)}{\overline{\lambda}(\Delta)^2}$ and $b=\frac{\overline{\lambda}(\Delta)\epsilon \Delta^{2H}}{D_{H} S(\Delta,H,N)}$.
Here the function $\mathcal{H}(u)=1+u-\sqrt{1+2u} $ and $\bar{\lambda}(\Delta) = 2 \max\left\{  \lambda_j(\Delta) \right\}$, where 
$\lambda_j(\Delta)$ ($j=1,2,...,N-\Delta$) are the eigenvalues of the $(N-\Delta) \times (N-\Delta)$ positive-definite covariance matrix  ${\Sigma}(\Delta)$ for
the vector of increments for FBM.  Moreover the $S(\Delta,H,N)$ function is defined as
\begin{equation}
S(\Delta,H,N)=\sum_{i=0}^{N-\Delta-1}[(i+\Delta)^{2H}-2i^{2H}+|i-\Delta|^{2H}]^2.
\end{equation}
It is worthwhile noting that for the FBM case the eigenvalues of the covariance matrix ${\Sigma}(\Delta)$ are not given in explicit form and need to be calculated numerically. Also note that Eq. (\ref{eq-dev-tamsd-fbm}) is independent of the generalized diffusion coefficient $D_H$ which gets canceled both in $a$ and $b$.

\section{Connection between Extreme value statistic and Large deviation theory}
\label{connect}

Consider $M$ discrete trajectories, $\{\{X_1,X_2,....,X_N\}_1$, $\{X_1,X_2,....,X_N
\}_2$, $\ldots,,\{X_1,X_2,....,X_N\}_M\}$ of length $N$ of a given process. Let $Y_j$
be a statistic over each trajectory $j$, $j\in\{1,2,...,M\}$ (for instance, $Y$ could
be the TAMSD). Large deviation theory deals with the probability that $P(Y>\epsilon)
\leq \exp(-I)$, where $I$ is the rate function and $\epsilon$ is the deviation
parameter. On the other hand, the extreme value statistic deals with the probability
$P\left(\max\{Y_1,Y_2,..,Y_M\}>z\right)$. This probability can be written as
\begin{eqnarray*}
P\left(\max\{Y_1,Y_2,..,Y_M\}>z\right)&&\\
&&\hspace*{-3.2cm}=1-P\left(\max\{Y_1,Y_2,..,Y_M\}\leq z\right)\\
&&\hspace*{-3.2cm}=1-P(Y_1 \leq z,Y_2 \leq z,...,Y_M \leq z) \\
&&\hspace*{-3.2cm}=1-\prod_{j=1}^{M}P(Y_{j} \leq z) \\
&&\hspace*{-3.2cm}=1-P^M(Y_1\leq z)\\
&&\hspace*{-3.2cm}=1-[1-P(Y_1>z)]^M.
\end{eqnarray*}
The last three equalities come from the fact that the considered trajectories represent independent realizations of the same process.

\section{Simulated processes}
\label{simproc}

For our analysis in the central Fig.~\ref{fig1} we simulate 100 trajectories
each for different processes. The number of trajectories is of the same order
as in the experimental datasets we analyze.\\

\textit{Brownian Motion} (BM): Brownian motion is characterized by the Langevin equation in the overdamped limit as \cite{vankampen,gardiner}
\begin{equation}
 \label{bm_langevin}
\frac{dX(t)}{dt}=\sqrt{2D}\eta(t),
\end{equation}
driven by the white Gaussian noise $\eta(t)$ with zero mean and autocorrelation
function $\langle\eta(t_1)\eta(t_2)\rangle=\delta(t_1-t_2)$.  The parameter $D$ is the diffusion coefficient.\\

\textit{Fractional Brownian Motion} (FBM): Fractional Brownian motion has been used to explain anomalous diffusion in a number of experiments
\cite{hurs51,hurs65,weis09,wero09,klaf10,burn10,jeon11,weber10}, where the underlying process had long-range correlations. FBM \cite{mand82,mand68} is given by the Langevin equation 
\begin{equation}
 \label{fbm_langevin}
\frac{dX_{\mathrm{FBM}}(t)}{dt}=\eta_{\mathrm{H}}(t),
\end{equation}
driven by the fractional Gaussian noise (fGn) $\eta_{\mathrm{H}}(t)$ with autocorrelation
function 
\begin{equation}
\label{fGn_corel}
\langle\eta_\mathrm{H}(t_1)\eta_\mathrm{H}(t_2)\rangle=2H(2H-1)D_H\times|t_1-
t_2|^{2(H-1)},
\end {equation}
where $D_H$ is the generalized diffusion coefficient and $H$ is the Hurst index, which is related to the anomalous diffusion
exponent $\alpha$ as $H=\alpha/2$.  \\

\textit{Scaled Brownian Motion} (SBM): Scaled Brownian motion has been used as a model of anomalous diffusion in numerous experiments
\cite{weiss07,verk98,berl08,hoys06,dous92,boon13}, particularly those with fluorescence recovery after photobleaching \cite{saxt01}.
SBM \cite{metz14,sbmralf} is characterized by Eq.~(\ref{bm_langevin}) but with a time-dependent diffusivity given by $D(t)=D_0 t^{\alpha-1}$, 
with constant $D_0$ and the anomalous diffusion exponent $\alpha$. The parameter $0<\alpha<1$ leads to a subdiffusive MSD while $1<\alpha<2$ leads to a superdiffusive MSD.\\

\textit{Continuous Time Random Walk} (CTRW): The subdiffusive CTRW has been used to describe a number of experiments \cite{scher75,weitz04,chaikin11,krapf11,swinney93} exhibiting anomalous diffusion. 
It is a renewal process with 
Gaussian jumps with an asymptotic power-law distributed waiting time between successive jumps \cite{scher75,montroll69,metz14}. The asymptotic probability density function (PDF) of the waiting time $\tau$ is given by
$\psi(\tau)\approx \tau_0^{\alpha}\tau^{(-1-\alpha)}$, where $0<\alpha<1$ is the anomalous diffusion exponent of the MSD. We refer to
\cite{hans07} for details of the simulation.\\

\textit{Superstatistical process}: By a superstatistical process \cite{beck03,beck06} we mean a process which is defined by Eq.~(\ref{bm_langevin})
where the diffusion coefficient is a random variable, that is, there exists a distribution of diffusivities over the tracers in a single particle
tracking experiment. The convolution of such distributions of diffusivities with a Gaussian distribution can give
rise to non-Gaussian displacement distributions routinely observed in many experiments \cite{hapc09,gran12,gran14,schw15d,
rubi17,cher19,gupt18,tong16,java16}.
As in many of these experiments, the diffusivity has a Rayleigh-like distribution, for our simulated  superstatistical process
we applied the Rayleigh distribution for the diffusivity,
\begin{equation}
 p(D)=\frac{D}{D_{0}^2}\exp{\left(-\frac{D^2}{2D_{0}^2}\right)},
 \label{rayleigh}
\end{equation}
where $D_{0}$ is the scale parameter of the Rayleigh distribution and is related to the mean $\langle D\rangle=D_{0}\sqrt{(\pi/2)}$.\\

\textit{Diffusing Diffusivity} (DD): The minimal DD model can be expressed as the set of stochastic differential equations
\cite {seno17}
\begin{subequations}
\begin{eqnarray}
&&\dfrac{dX_{DD}(t)}{dt}=\sqrt{2D(t)}\eta_{1}(t),
\label{eqdda}\\
&&D(t)=Y^2(t),\label{eqddb}\\
&&\dfrac{dY(t)}{dt}=-\dfrac{Y(t)}{\tau_{c}}+\sigma\eta_{2}(t),\label{eqddc}
\end{eqnarray}
\label{eq-dd-3-set}
\end{subequations}\\
where the time dependent diffusion
coefficient is defined as the square of the Ornstein-Uhlenbeck process $Y(t)$ and $\tau_{c}$ is the relaxation time to the stationary limit
\cite{uhle30}. $\eta_{1}(t)$ and $\eta_{2}(t)$ are independent white Gaussian noise with zero mean and unit variance. In the long time, stationary limit the
diffusion coefficients are distributed roughly
exponentially \cite{seno17},
\begin{equation}
p(D)=(\pi D D_\star)^{-1/2}\exp[-D/D_\star],
\label{eq-final-expon-d-distr}
\end{equation}
where $D_\star=\sigma^2\tau_{c}$. The TAMSD for this DD model grows linearly with lag time but the PDF of the process is non-Gaussian (Laplacian) for times
less than the relaxation time $\tau_{c}$, and it crosses over to a Gaussian PDF for $t\gg\tau_{c}$. This behavior was seen in a number of experiments
\cite{gran12,gran14}.

\section{The Jarque-Bera test for Gaussianity}
\label{sec-JB}

In statistics, the Jarque-Bera (JB) test is a goodness-of-fit test used to recognize if the sample data have the skewness and kurtosis matching the Gaussian distribution. The test statistic is always nonnegative. If it is far from zero, then we can suspect, the data are not from the Gaussian distribution. The 
JB statistic for a random sample $x_1,x_2,...,x_n$ is defined as follows \cite{jb},
\begin{eqnarray}
JB=\frac{n}{6}\left(S^2+\frac{1}{4}(K-3)^2\right),
\end{eqnarray}
where $S$ and $K$ are the empirical skewness and kurtosis, respectively.

In the literature, the JB test based on the JB statistic is considered as one of the most effective tests for Gaussianity. It is especially useful in the problem of recognition between heavy- and light-tailed (Gaussian) distributions of the data.

\section{Supplementary figures}

We here present additional figures that we refer to in the main text.

\begin{figure*}
\includegraphics[scale=0.301,angle=270]{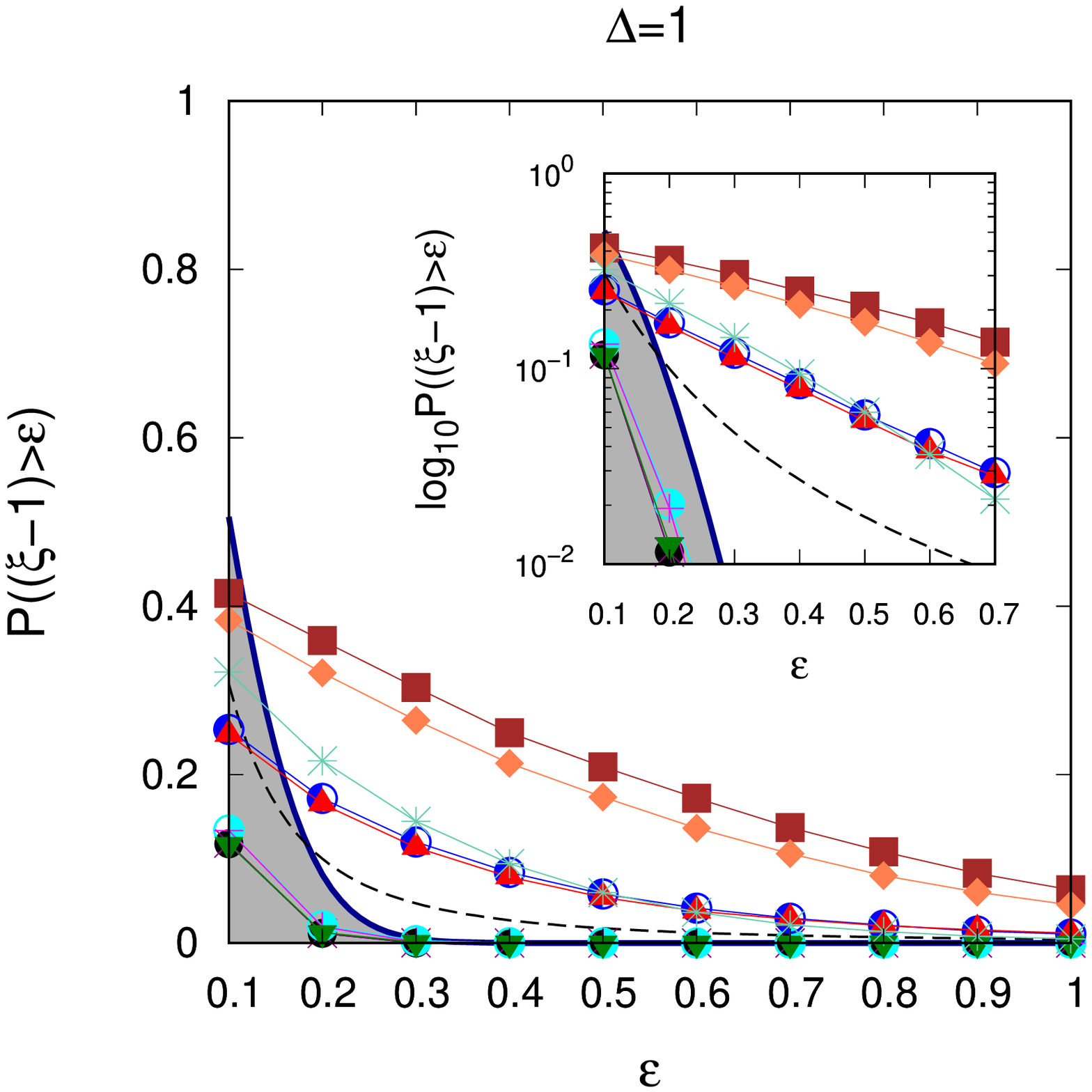}
\includegraphics[scale=0.301,angle=270]{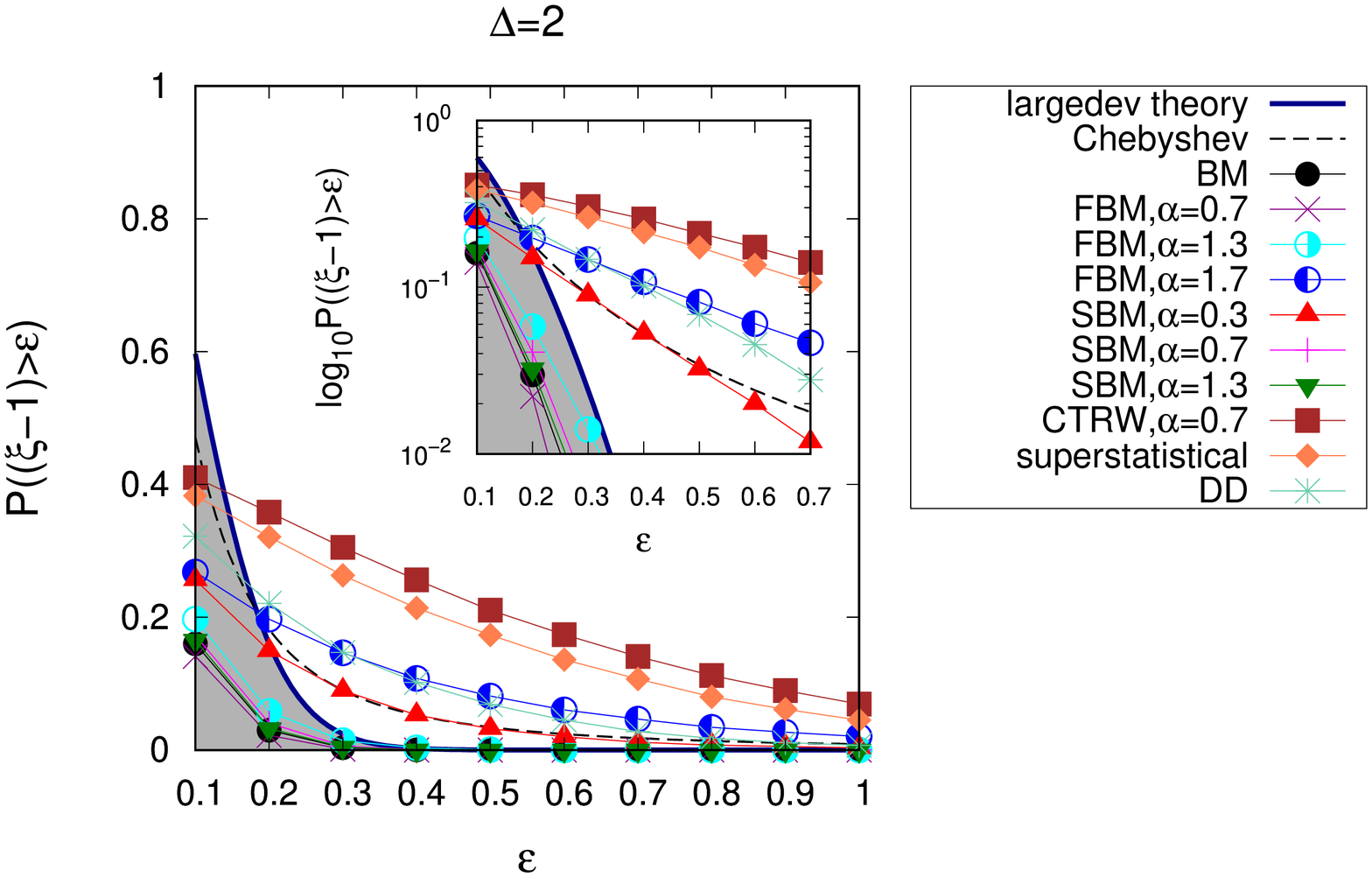}\\
\includegraphics[scale=0.301,angle=270]{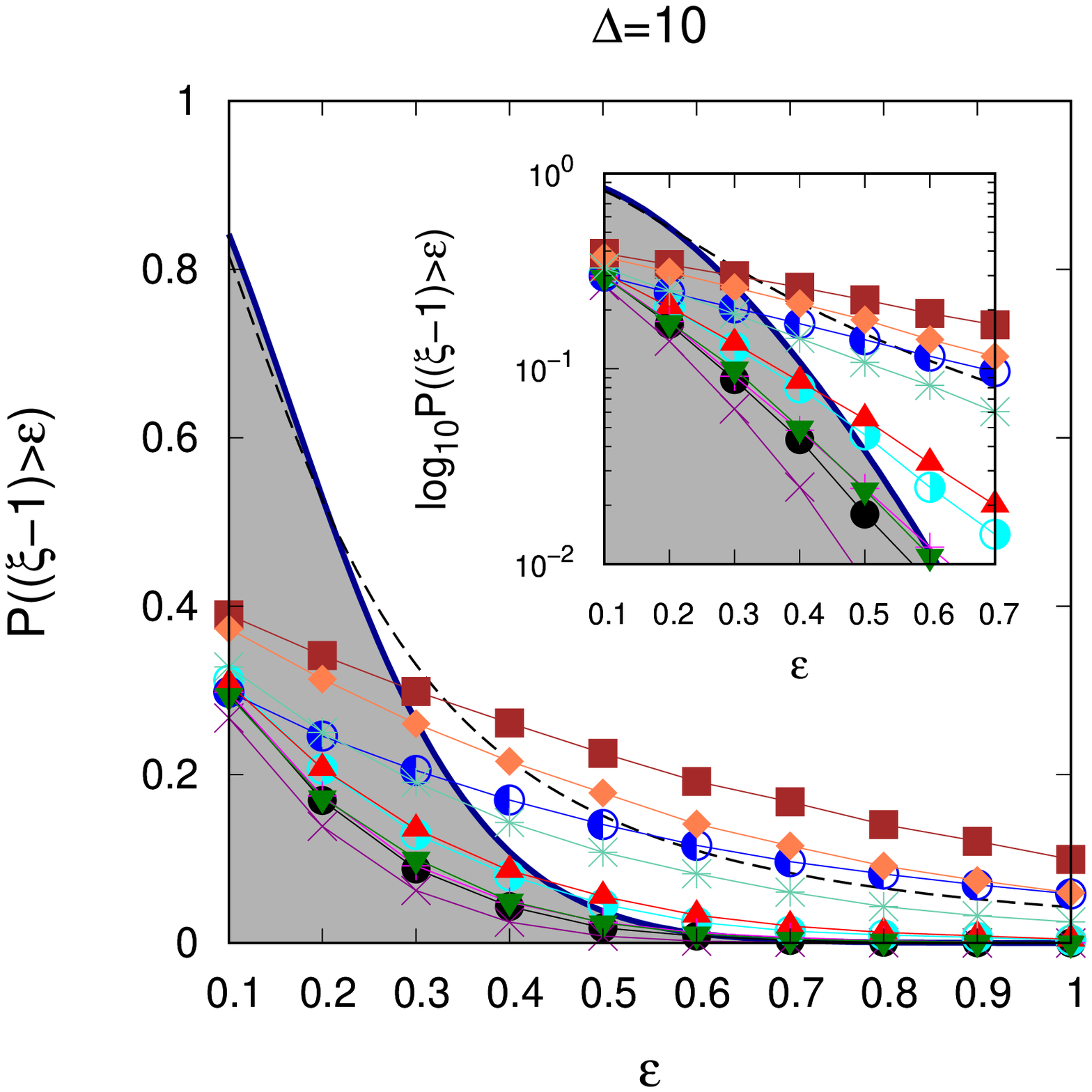}
\includegraphics[scale=0.301,angle=270]{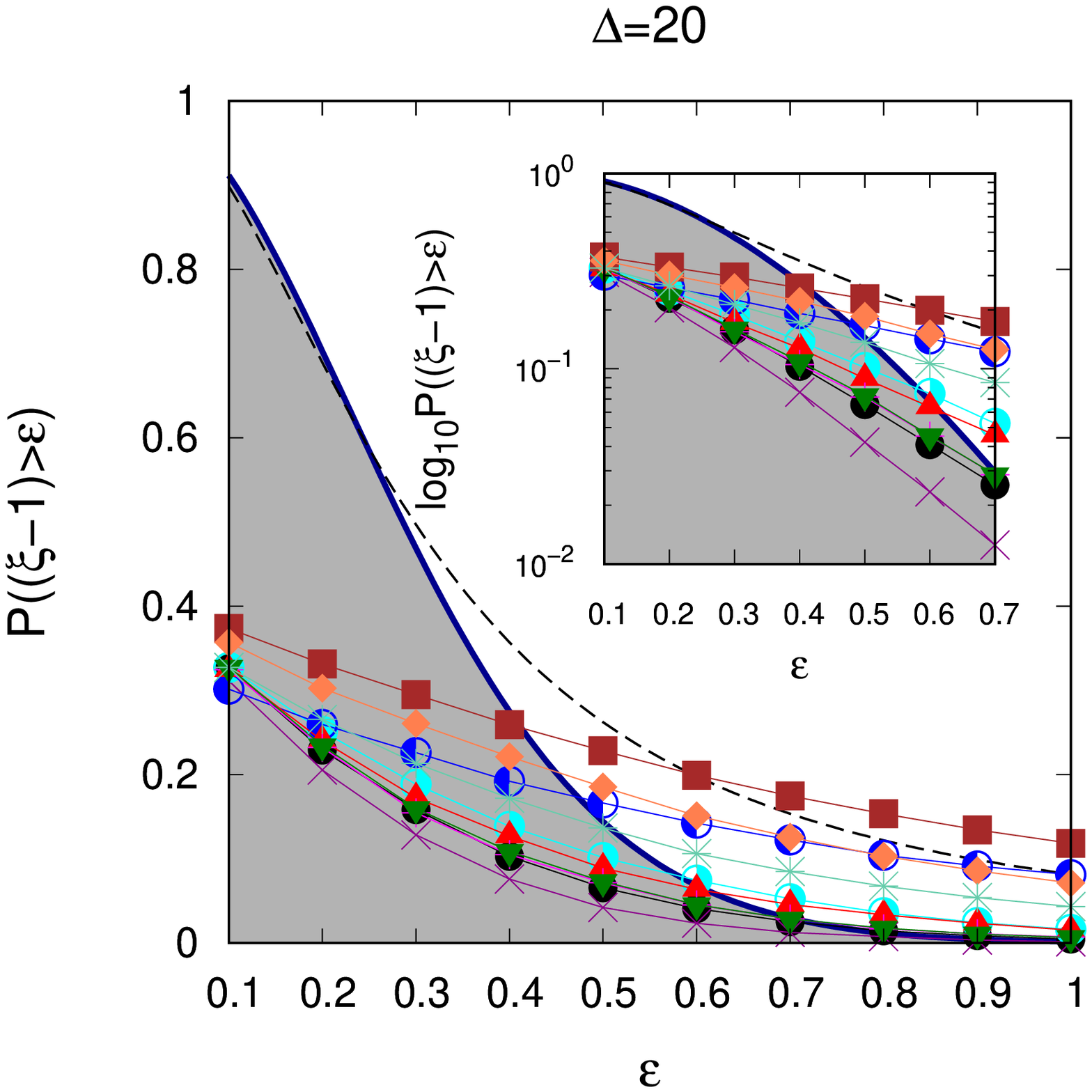}
\caption{Variation of the estimates of $P\left((\xi-1)>\epsilon\right)$ with
respect to $\epsilon$, for different simulated datasets with $N=300$, $M=10000$
and different values of $\Delta$.}
\label{figs1}
\end{figure*}

\begin{figure*}
\includegraphics[scale=0.301,angle=270]{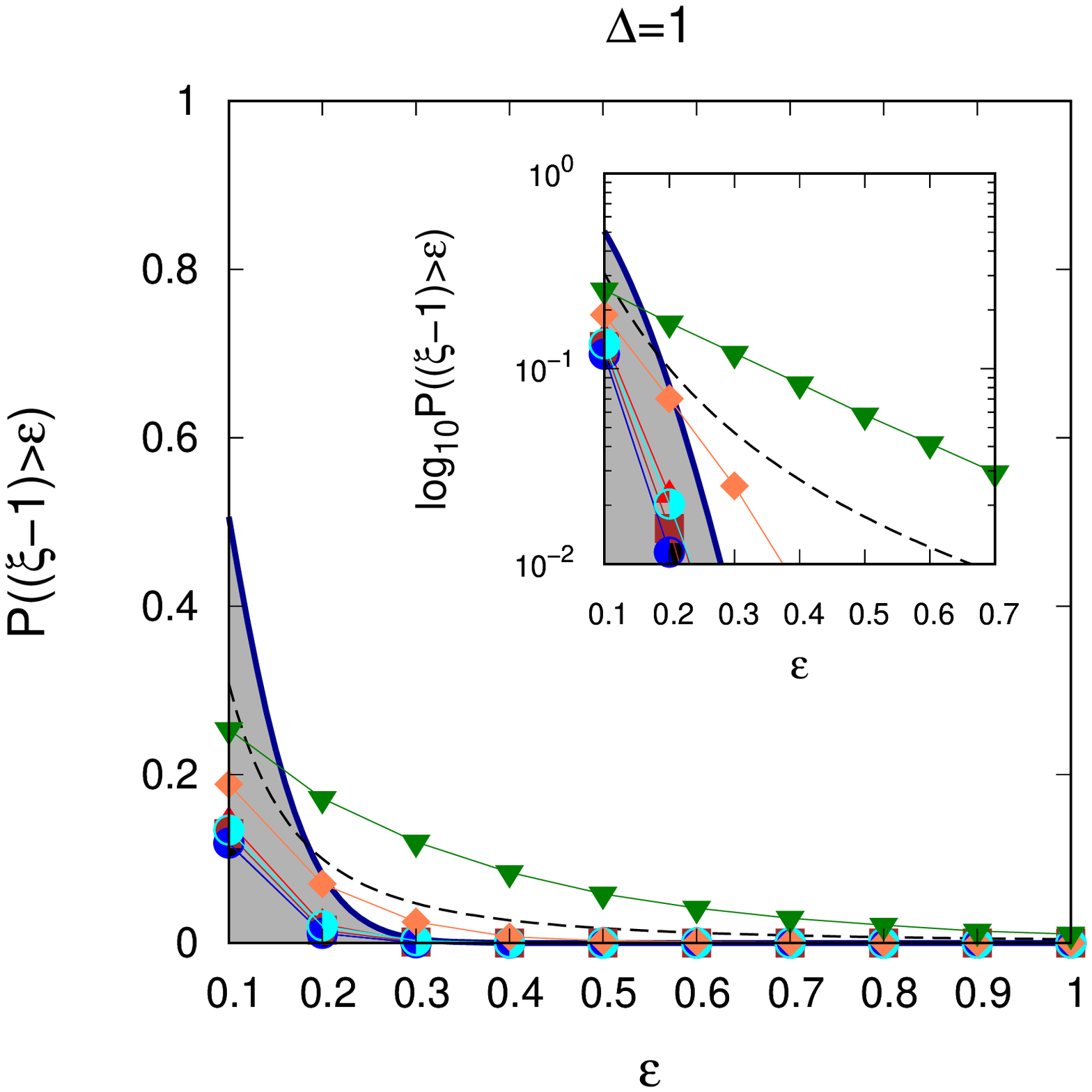}
\includegraphics[scale=0.301,angle=270]{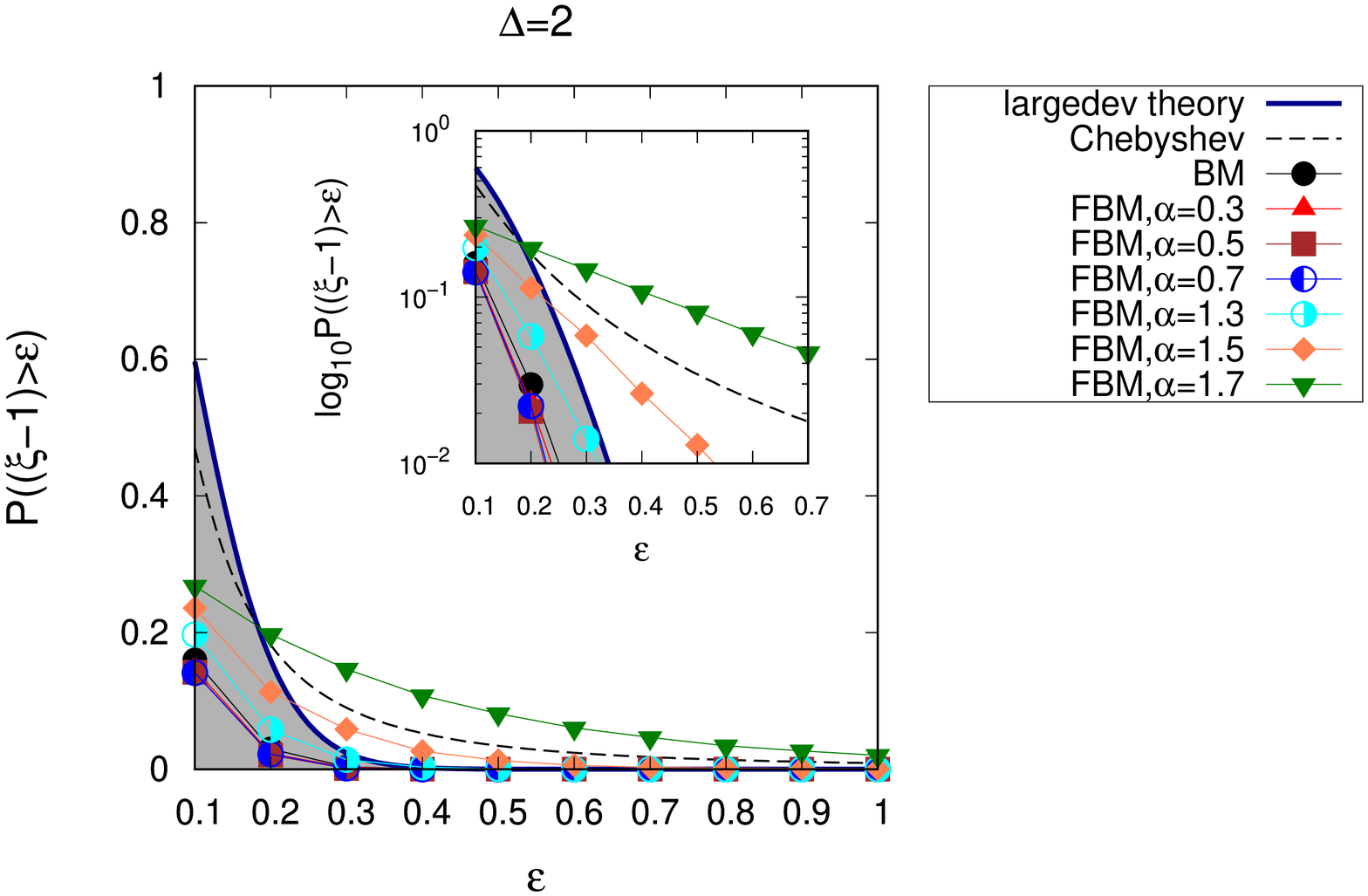}\\
\includegraphics[scale=0.301,angle=270]{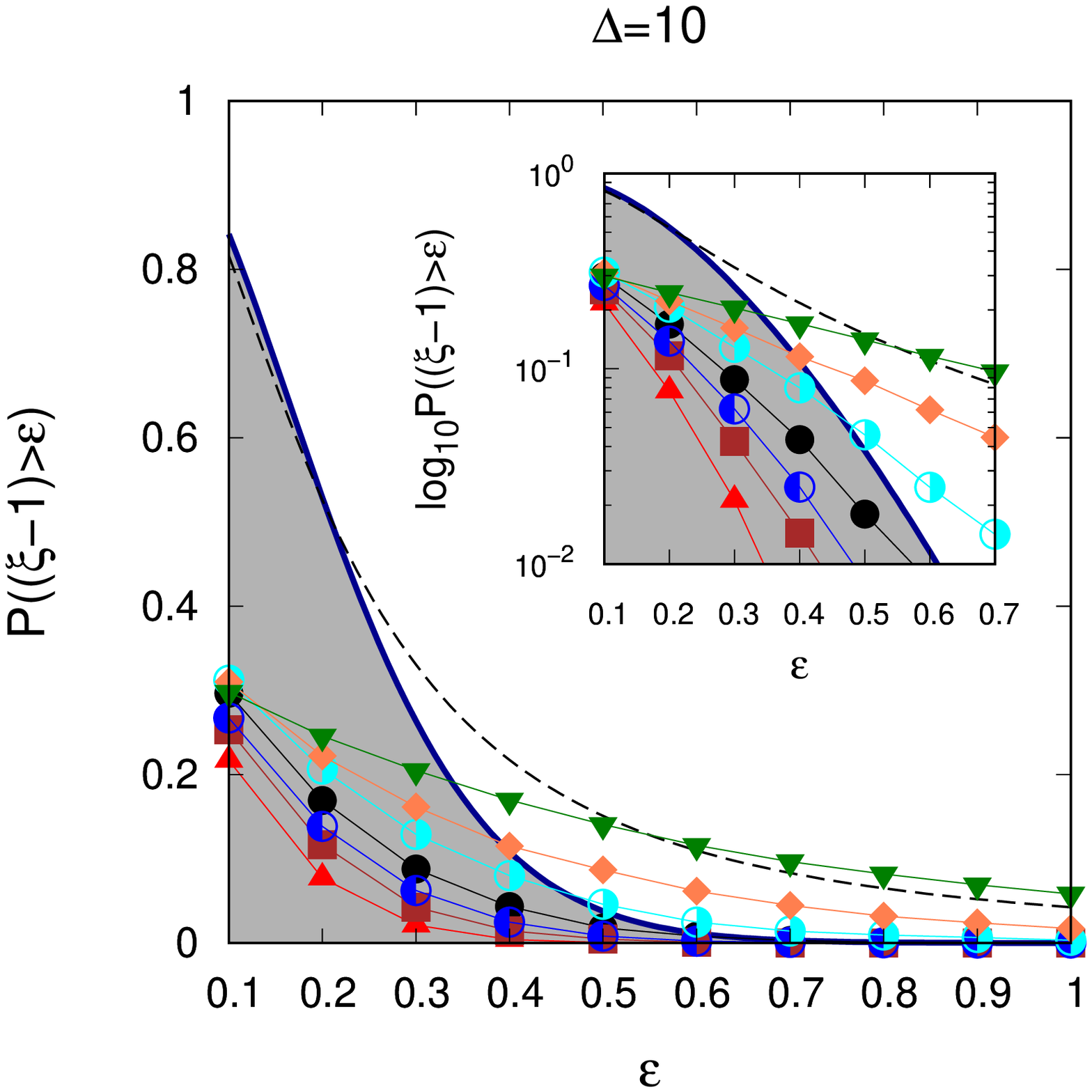}
\includegraphics[scale=0.301,angle=270]{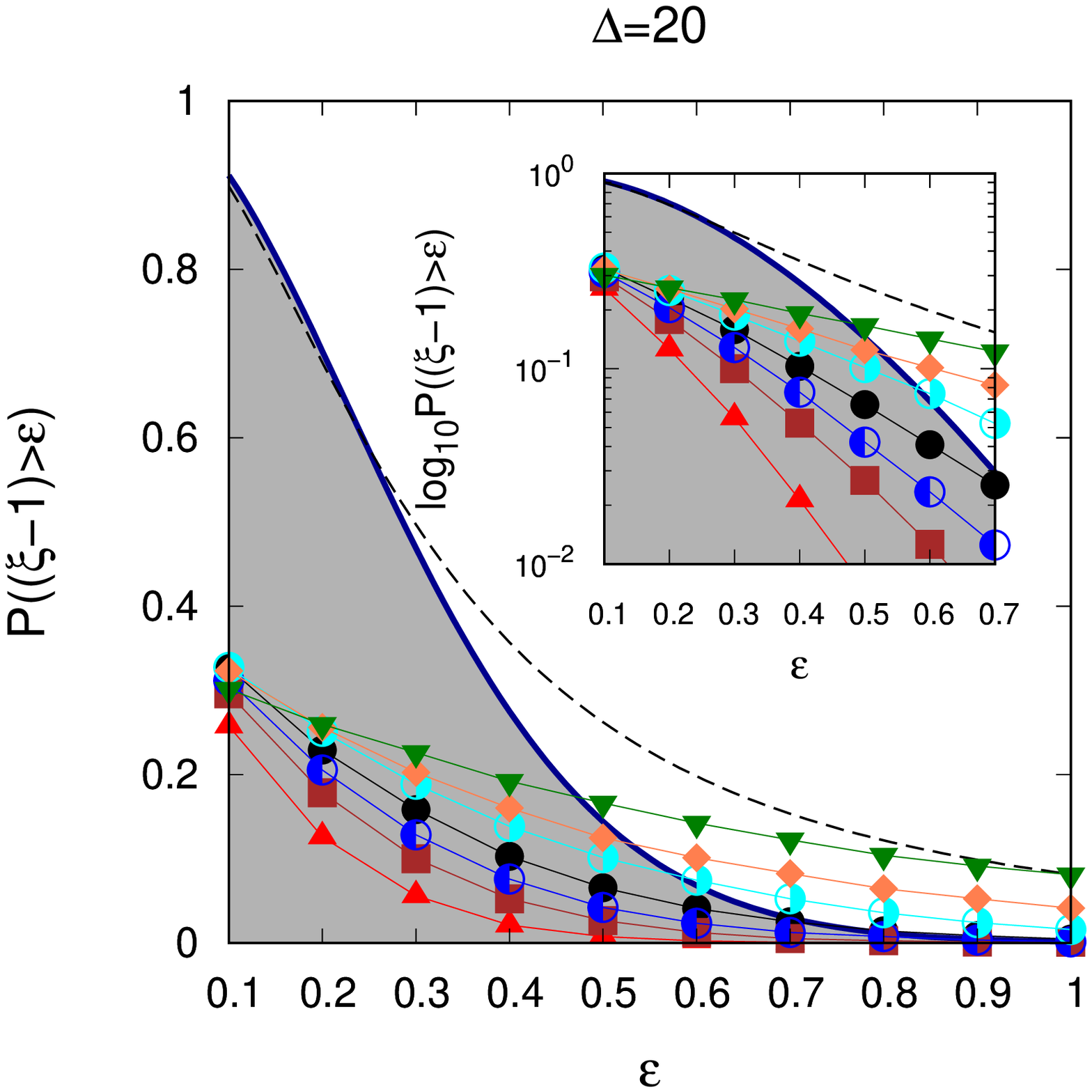}
\caption{Variation of the estimates of $P\left((\xi-1)>\epsilon\right)$ with
respect to $\epsilon$, for simulated FBM datasets of different anomalous
diffusion exponent, with $N=300$, $M=10000$ and different values of $\Delta$.
Superdiffusive FBM with large values of the anomalous diffusion exponent
$\alpha$ fall out of the Brownian domain (gray-shaded region in the plot)
for large values of the deviation parameter $\epsilon$.}
\label{figs2}
\end{figure*}

\begin{figure*}
\includegraphics[scale=0.301,angle=270]{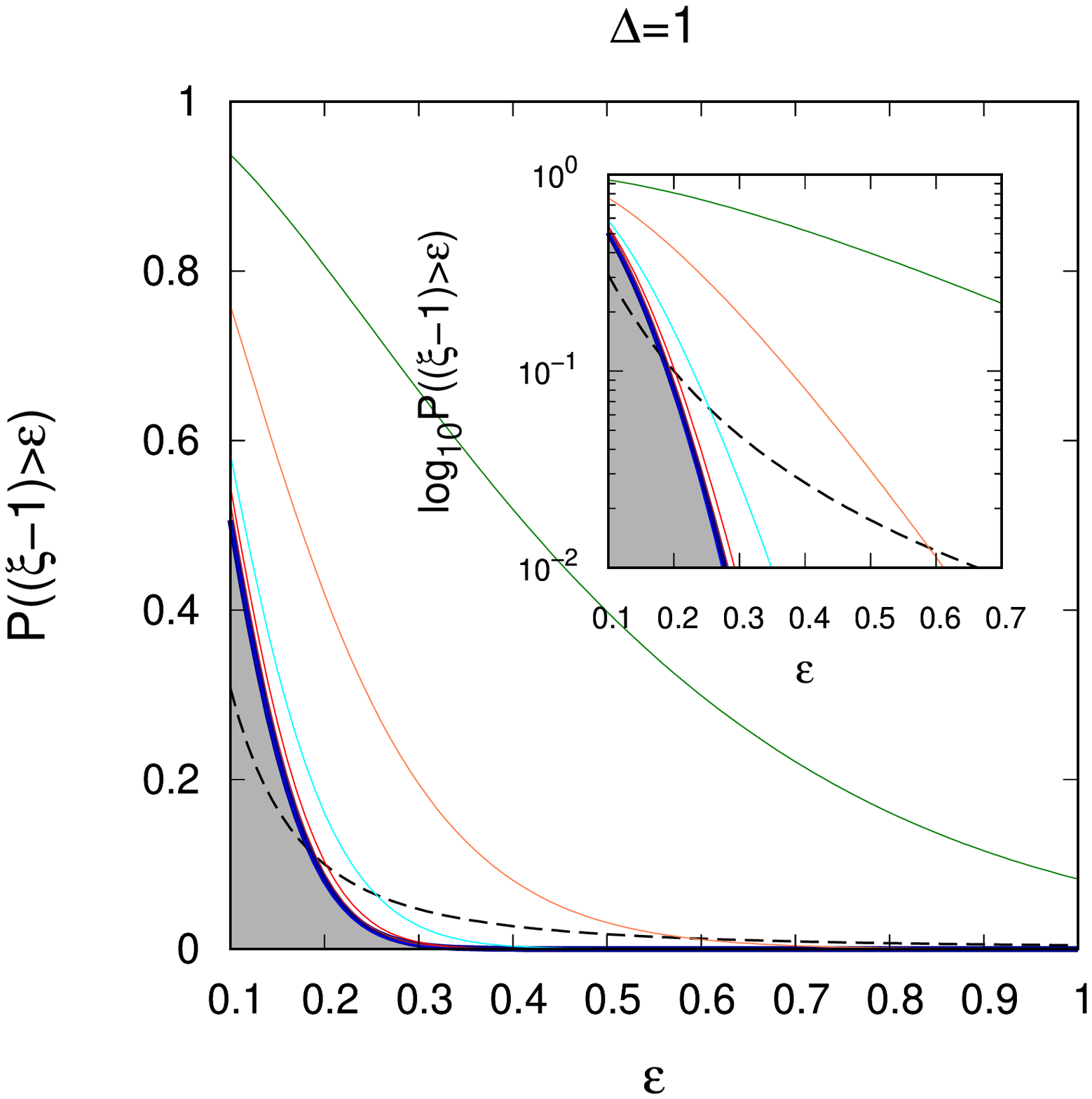}
\includegraphics[scale=0.301,angle=270]{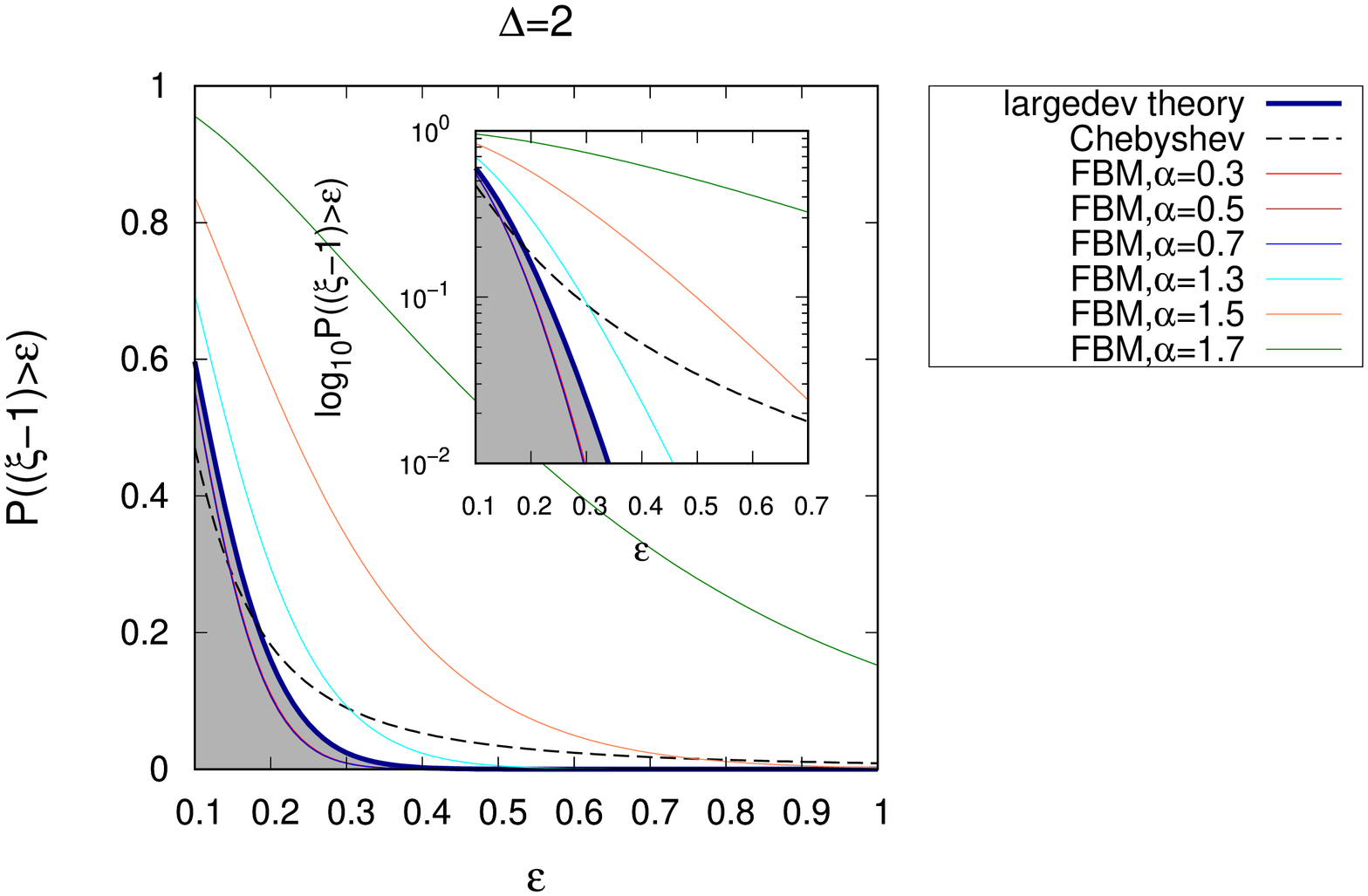}\\
\includegraphics[scale=0.301,angle=270]{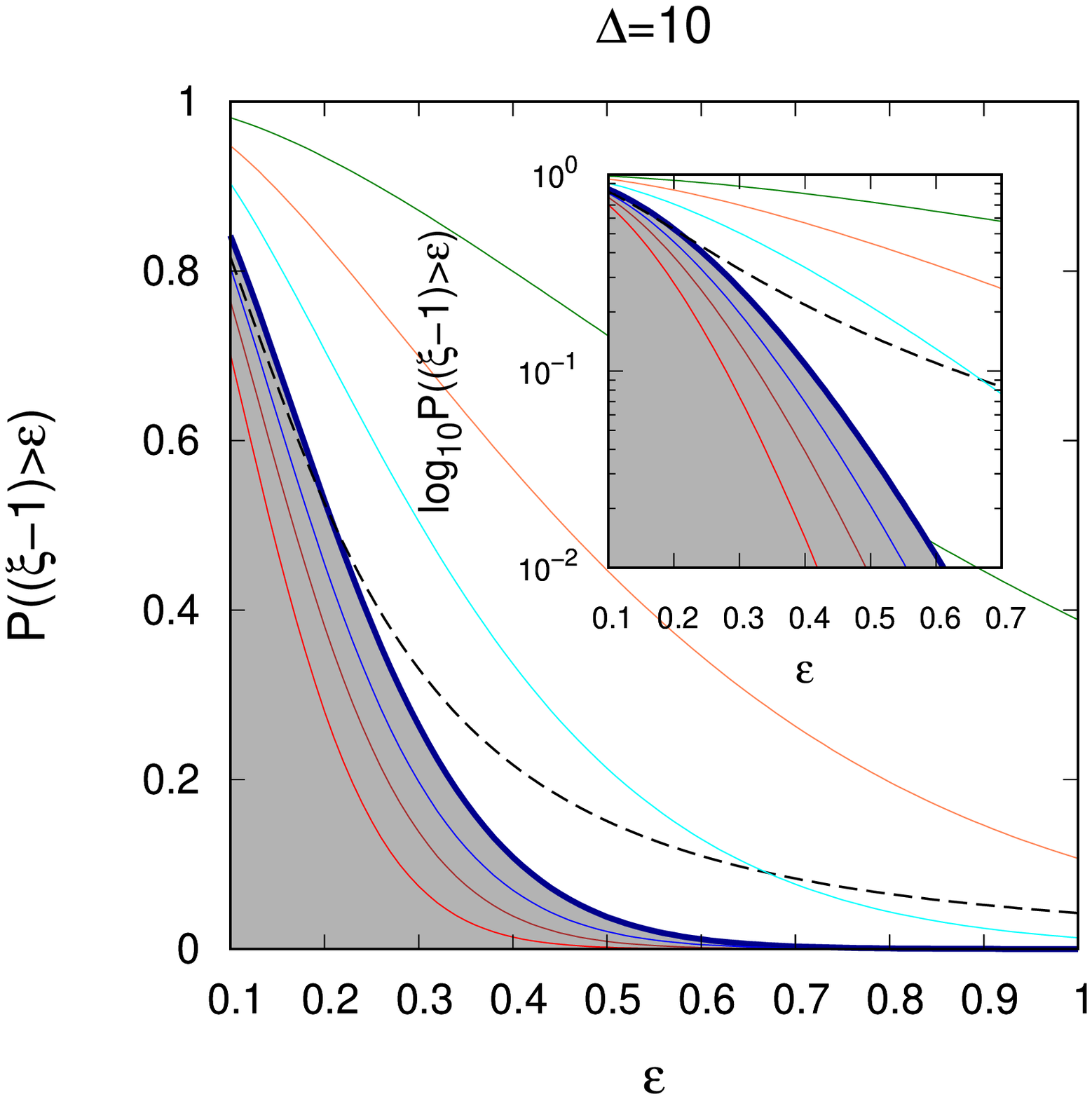}
\includegraphics[scale=0.301,angle=270]{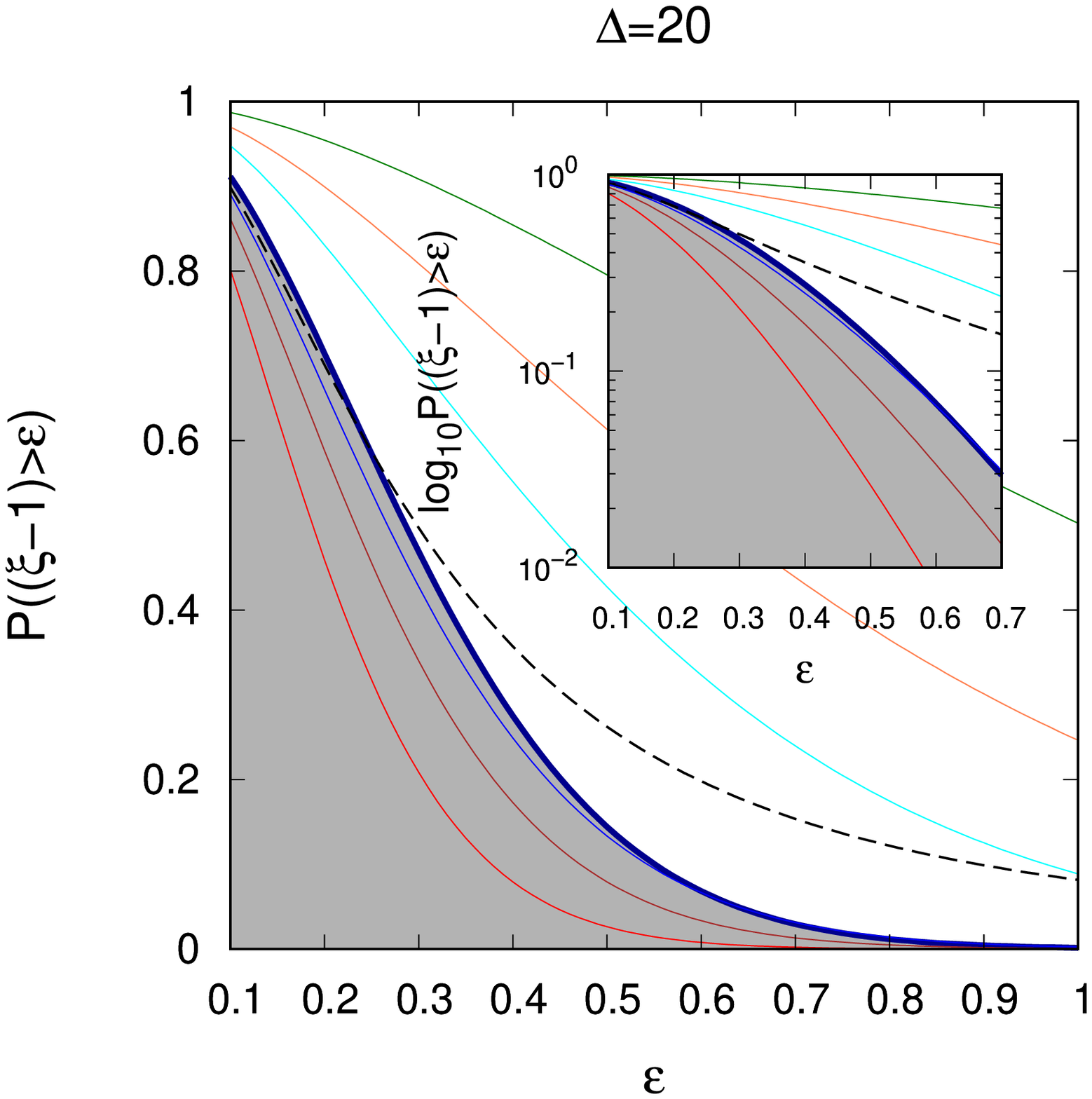}
\caption{Comparison of the theoretical curves for the variation of $P\left((\xi-1)
>\epsilon\right)$ with respect to $\epsilon$, for BM (labeled "largedev theory")
and FBM with different anomalous diffusion exponent. The curve labeled "Chebyshev"
is the theoretical curve of the Chebyshev inequality for BM. The different
sub-figures are for $N=300$ and different values of the lag time, $\Delta$.
Super-diffusive FBM with large values of the anomalous diffusion exponent $\alpha$
fall out of the Brownian domain (gray-shaded region in the plot), i.e. the
theoretical bounds on the deviations of the normalized TAMSD for superdiffusive
FBM are increasingly larger for larger values of $\alpha$. This supports the
simulation results in Fig.\ref{figs2}. No clear trend is observed for
subdiffusive FBM at short lag times, although a trend similar to superdiffusive
FBM appears at long lag times. This is, however, consistent with
Fig.~\ref{figs2}, where a clear trend for subdiffusive FBM is observed for
the simulated datasets only at long lag times.}
\label{figs3}
\end{figure*}

\begin{figure*}
\includegraphics[scale=0.301,angle=270]{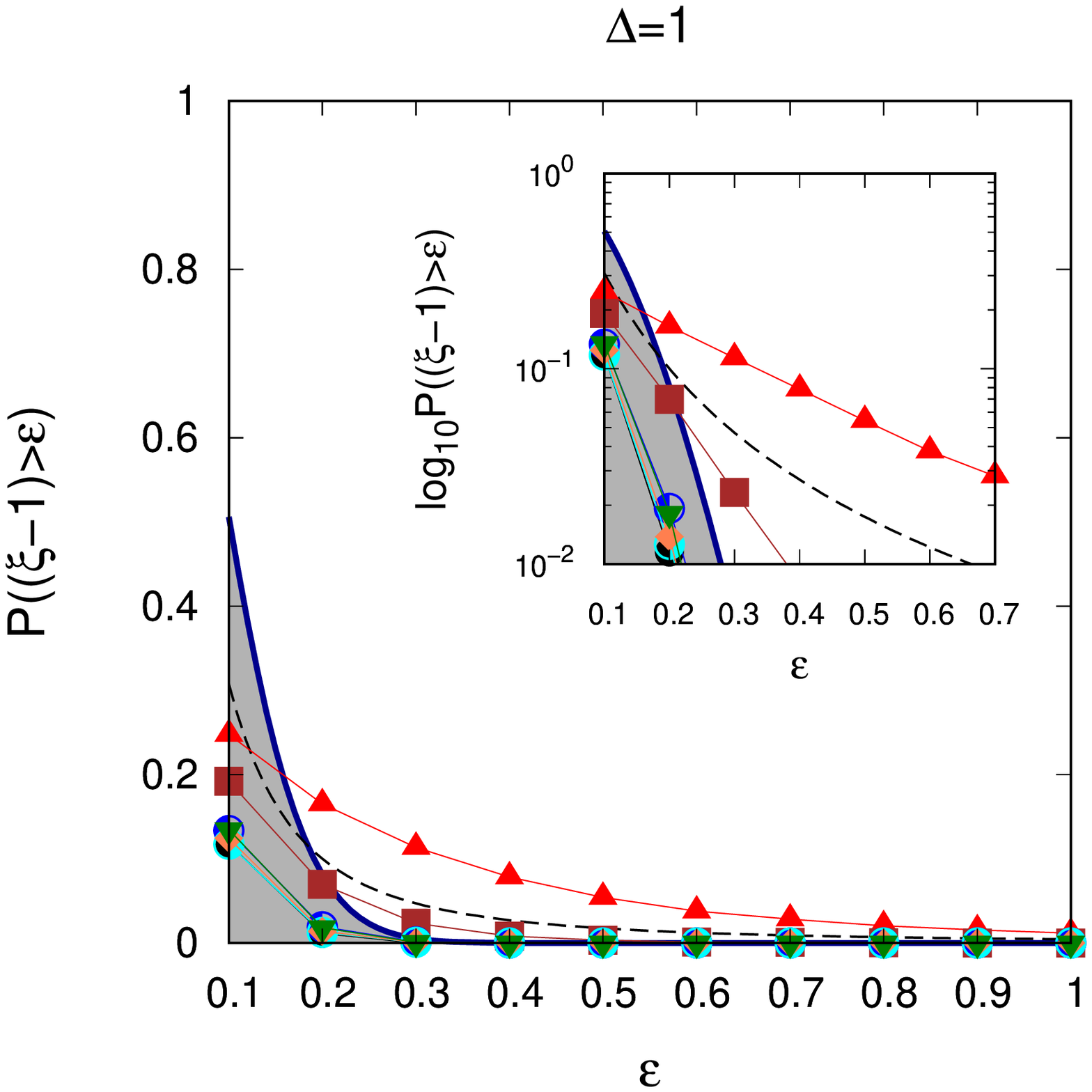}
\includegraphics[scale=0.301,angle=270]{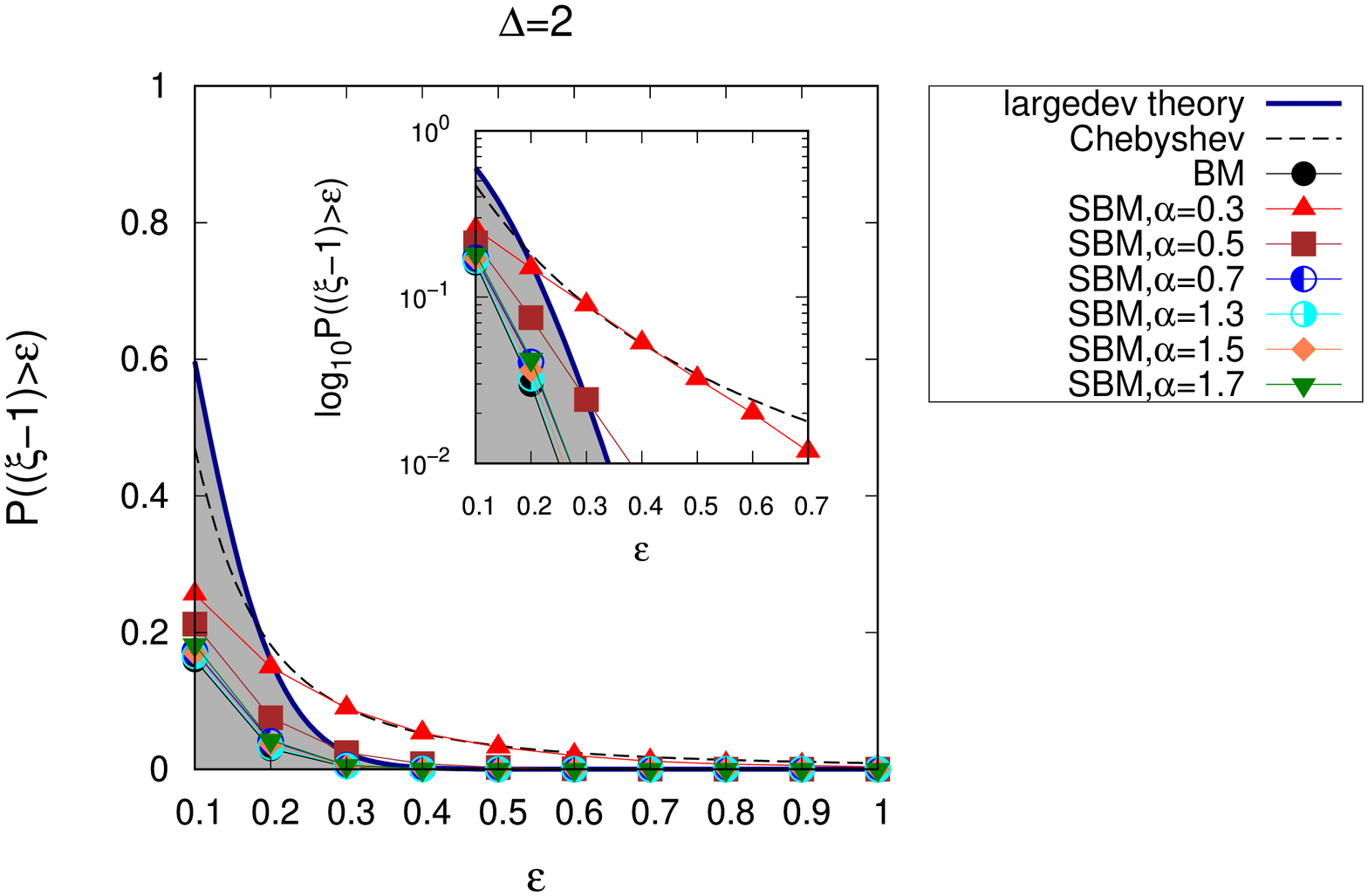}\\
\includegraphics[scale=0.301,angle=270]{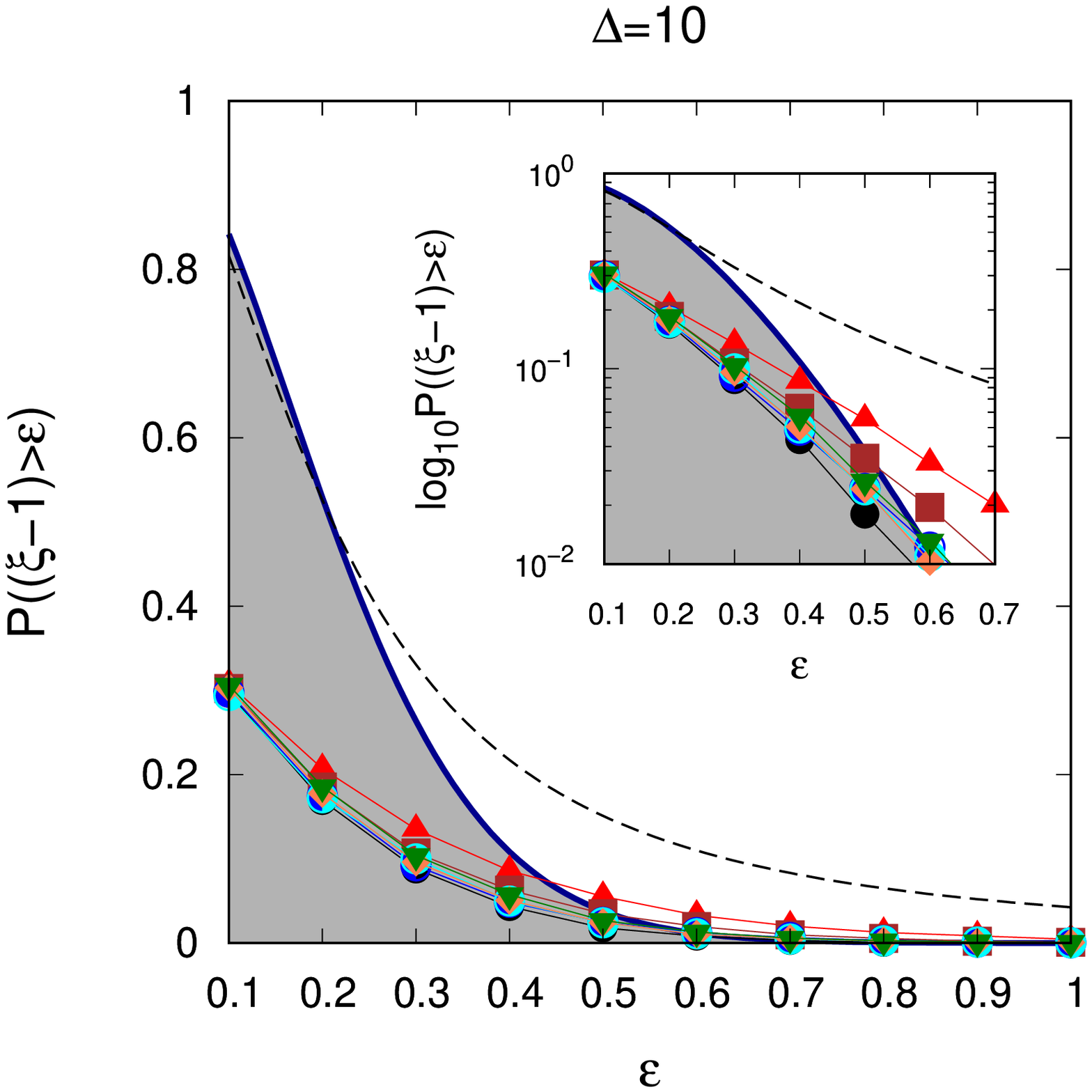}
\includegraphics[scale=0.301,angle=270]{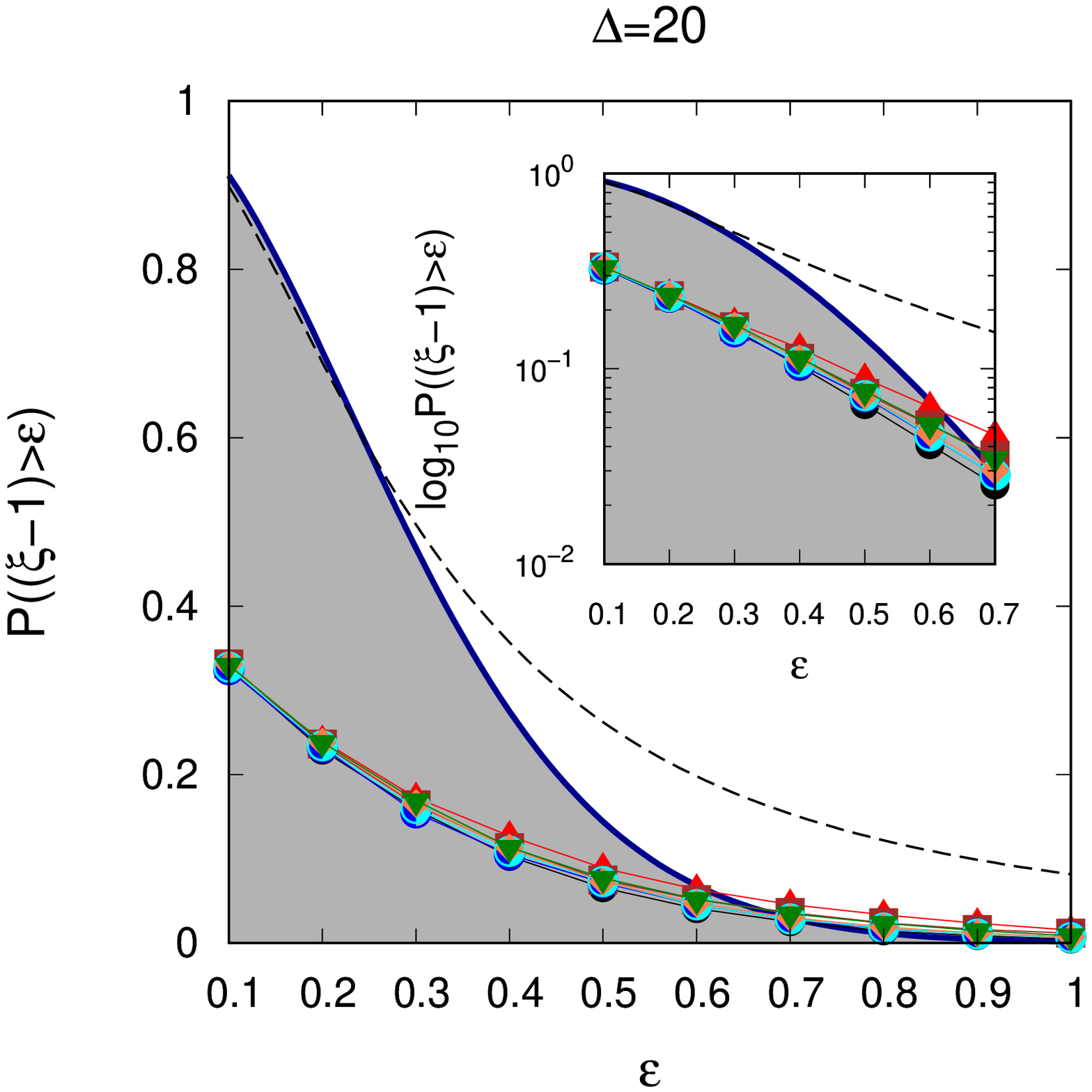}
\caption{Variation of the estimates of $P\left((\xi-1)>\epsilon\right)$ with
respect to $\epsilon$, for simulated SBM datasets of different anomalous
diffusion exponent, with $N=300$, $M=10000$ and different values of $\Delta$.
Subdiffusive SBM with small values of the anomalous diffusion exponent $\alpha$
fall out of the Brownian domain (gray-shaded region in the plot) for large values
of the deviation parameter $\epsilon$, significantly at short lag times $\Delta$.
This is in agreement with the behavior of the EB parameter reported for SBM in
Ref.~\cite{hadiseh}, namely that for subdiffusive SBM the EB parameter (or
equivalently the variance of the TAMSD) is larger for smaller values of the
anomalous diffusion exponent. Moreover, for subdiffusive SBM, it was also reported
in Ref.~\cite{hadiseh} that the EB parameter is larger for short lag times at fixed
values of the anomalous diffusion exponent. This explains why subdiffusive SBM can
be better distinguished from BM at small values of the lag time.}
\label{figs4}
\end{figure*}

\begin{figure*}
\includegraphics[width=6.8cm,angle=270]{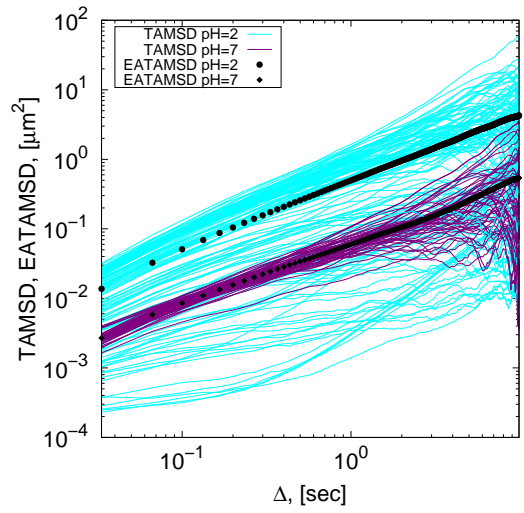}
\caption{TAMSD (and its ensemble average EATAMSD) plot considering 2-dimensional
motion of beads tracked in Mucin hydrogel at $pH=2$ and $pH=7$. There are $50$
trajectories at $pH=7$ and $131$ trajectories at $pH=2$. Each trajectory consists
of $N=300$. The estimated $\left<\beta\right>=\{1.09,0.94\}$ for $pH=\{2,7\}$
respectively \cite{mucin19}.}
\label{figs5}
\end{figure*}

\begin{figure*}
\includegraphics[width=6.8cm,angle=270]{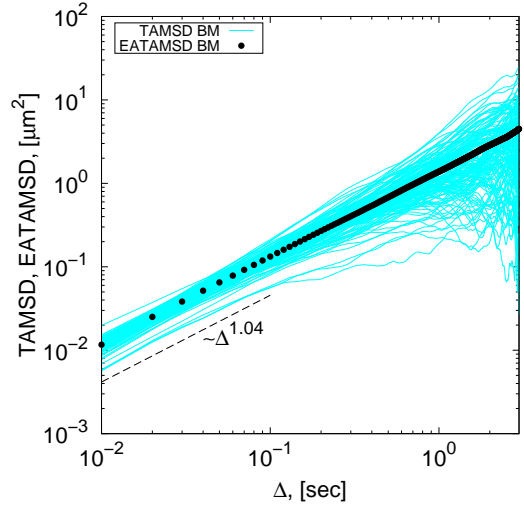}
\caption{TAMSD (and its ensemble average EATAMSD) plot considering 2-dimensional
motion of beads tracked in aqueous solution. There are $150$ trajectories with
$N=300$.}
\label{figs6}
\end{figure*}

\begin{figure*}
\includegraphics[width=6.8cm,angle=270]{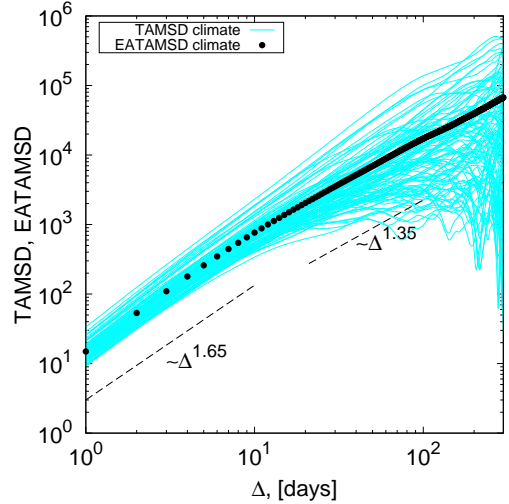}
\caption{TAMSD (and its ensemble average EATAMSD) plot of climate data after
taking cumulative sum of the temperature anomalies. There are $100$ trajectories
with $N=300$.}
\label{figs7}
\end{figure*}

\begin{figure*}
\includegraphics[scale=0.35,angle=270]{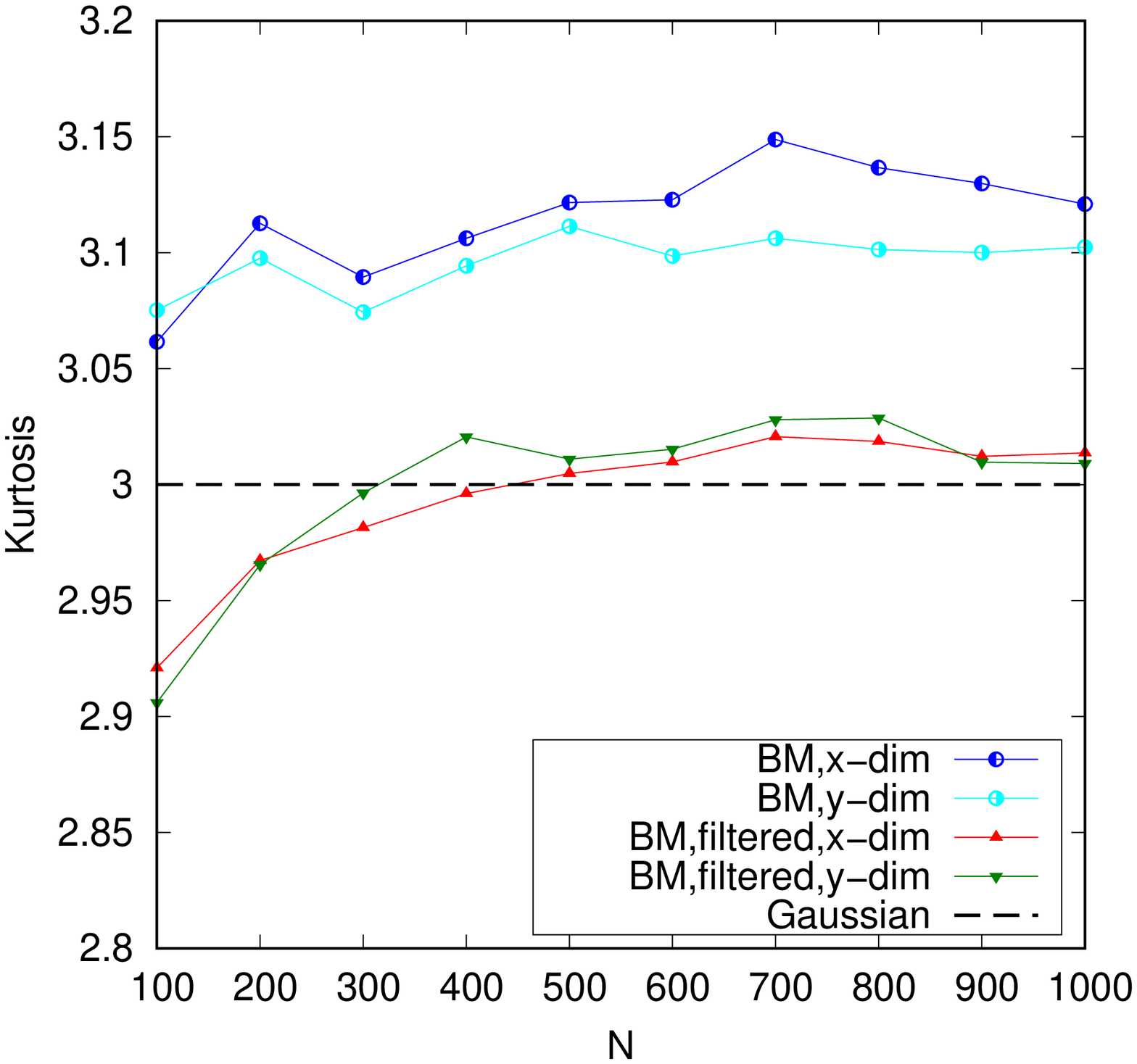}
\includegraphics[scale=0.35,angle=270]{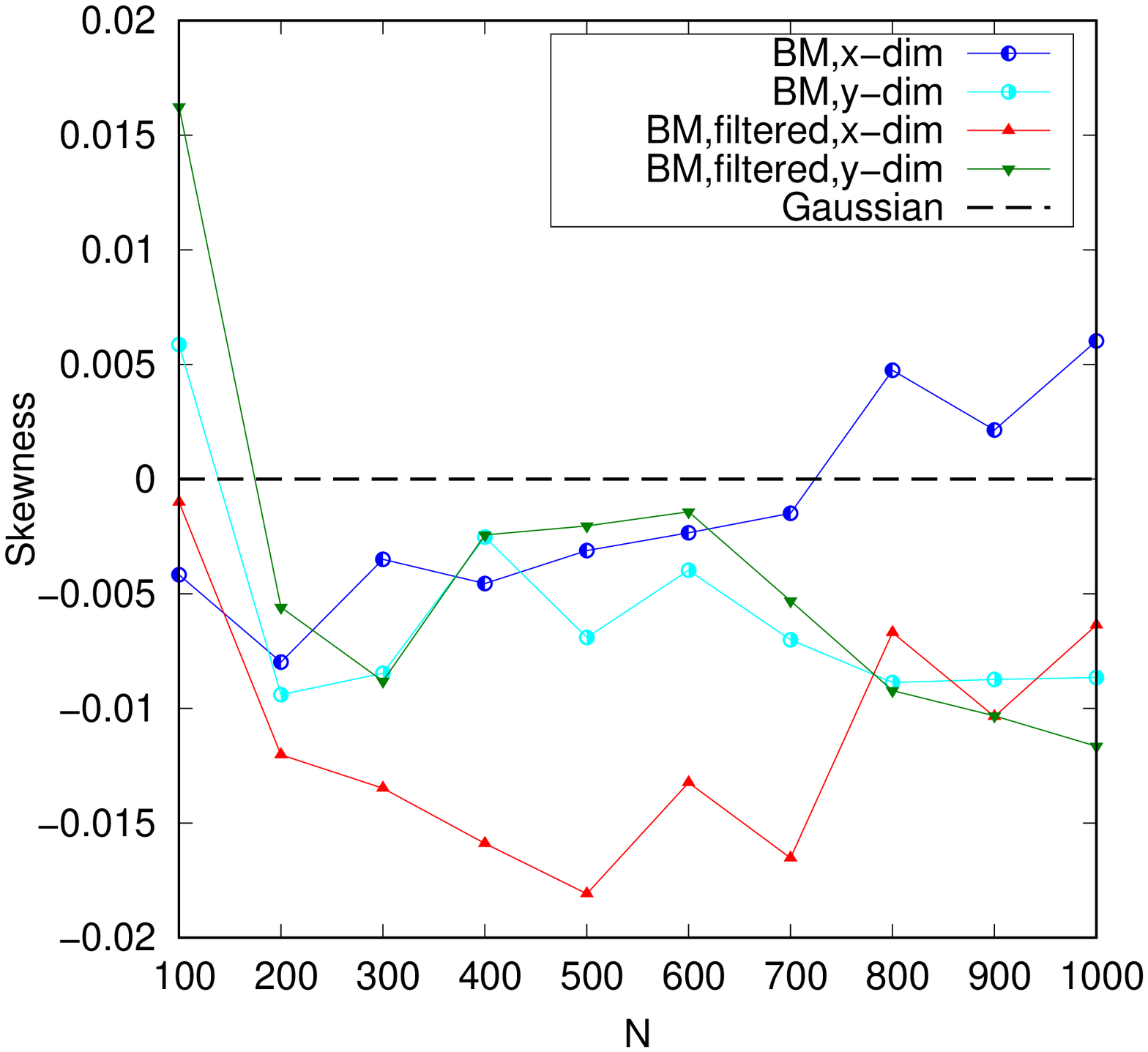}
\caption{Average (over the trajectories) kurtosis (left) and average skewness
(right) as function of $N$ for trajectories of polystyrene beads tracked in
aqueous solution. This plot shows how the set of trajectories filtered using
the JB test (labeled "BM,filtered,x-dim" and "BM,filtered,y-dim") shows
Gaussian behavior (particularly with respect to the kurtosis), as compared to
the full set of trajectories (labeled "BM,x-dim" and "BM,y-dim").}
\label{figs8}
\end{figure*}

\begin{figure*}
\includegraphics[scale=0.301,angle=270]{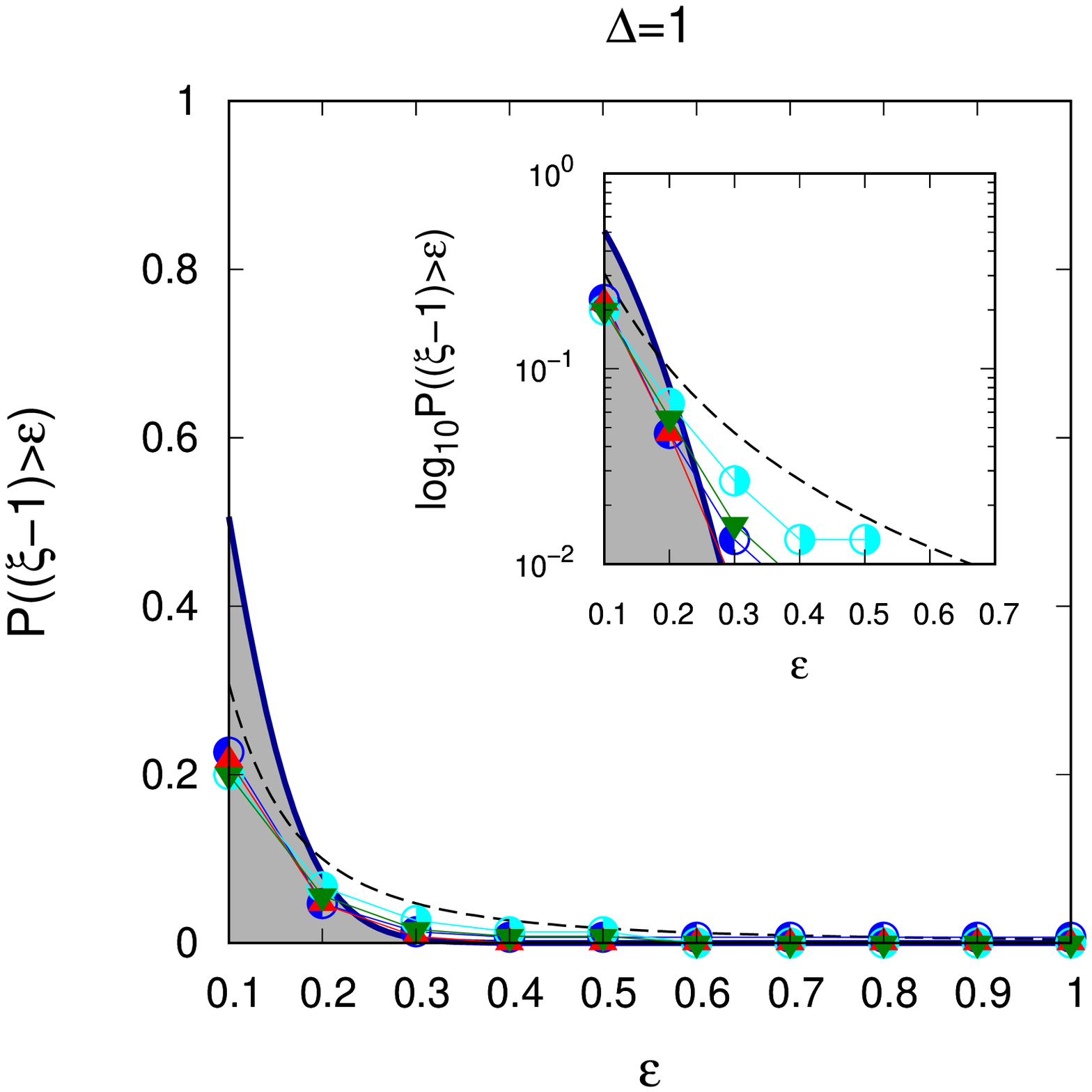}
\includegraphics[scale=0.301,angle=270]{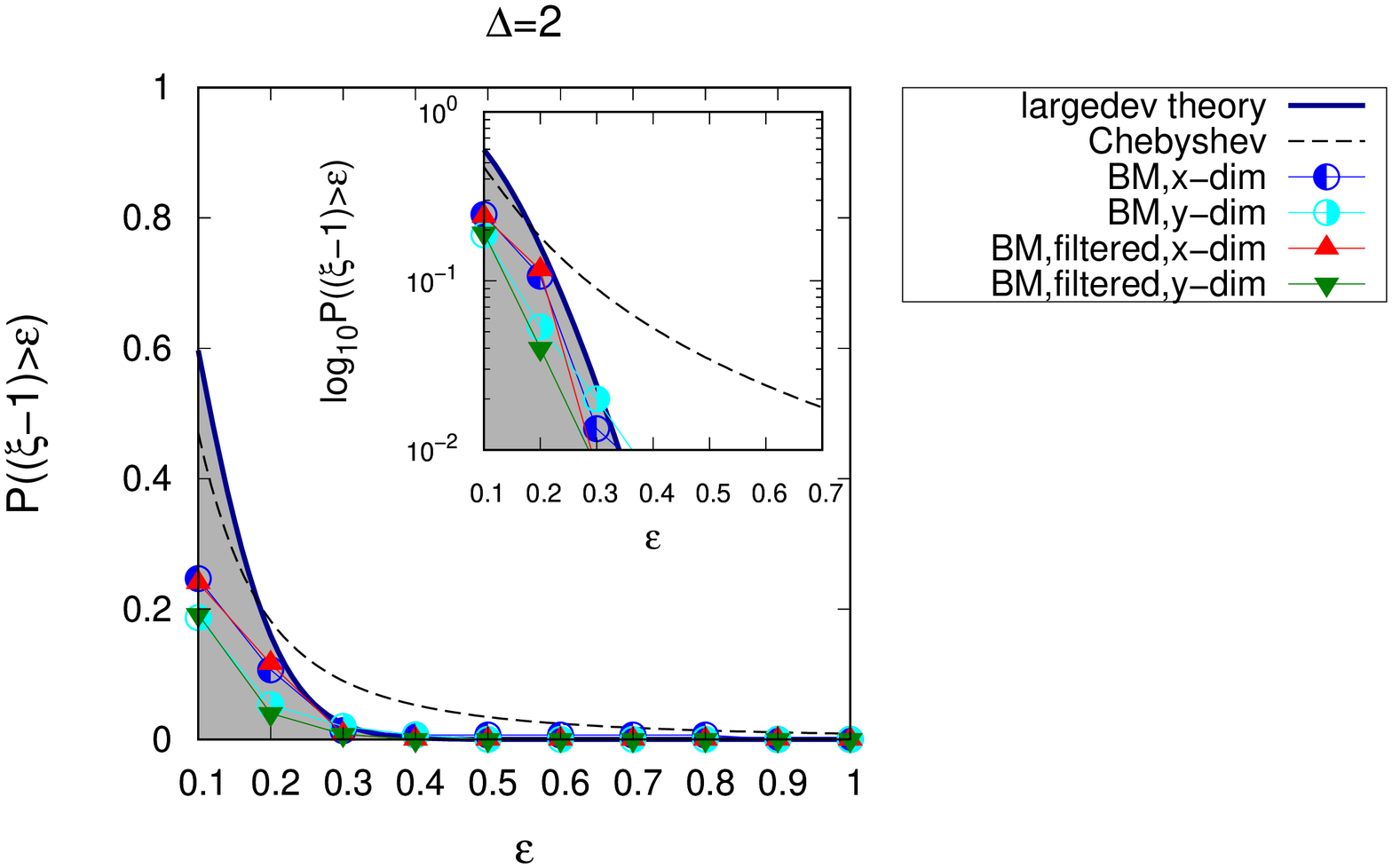}\\
\includegraphics[scale=0.301,angle=270]{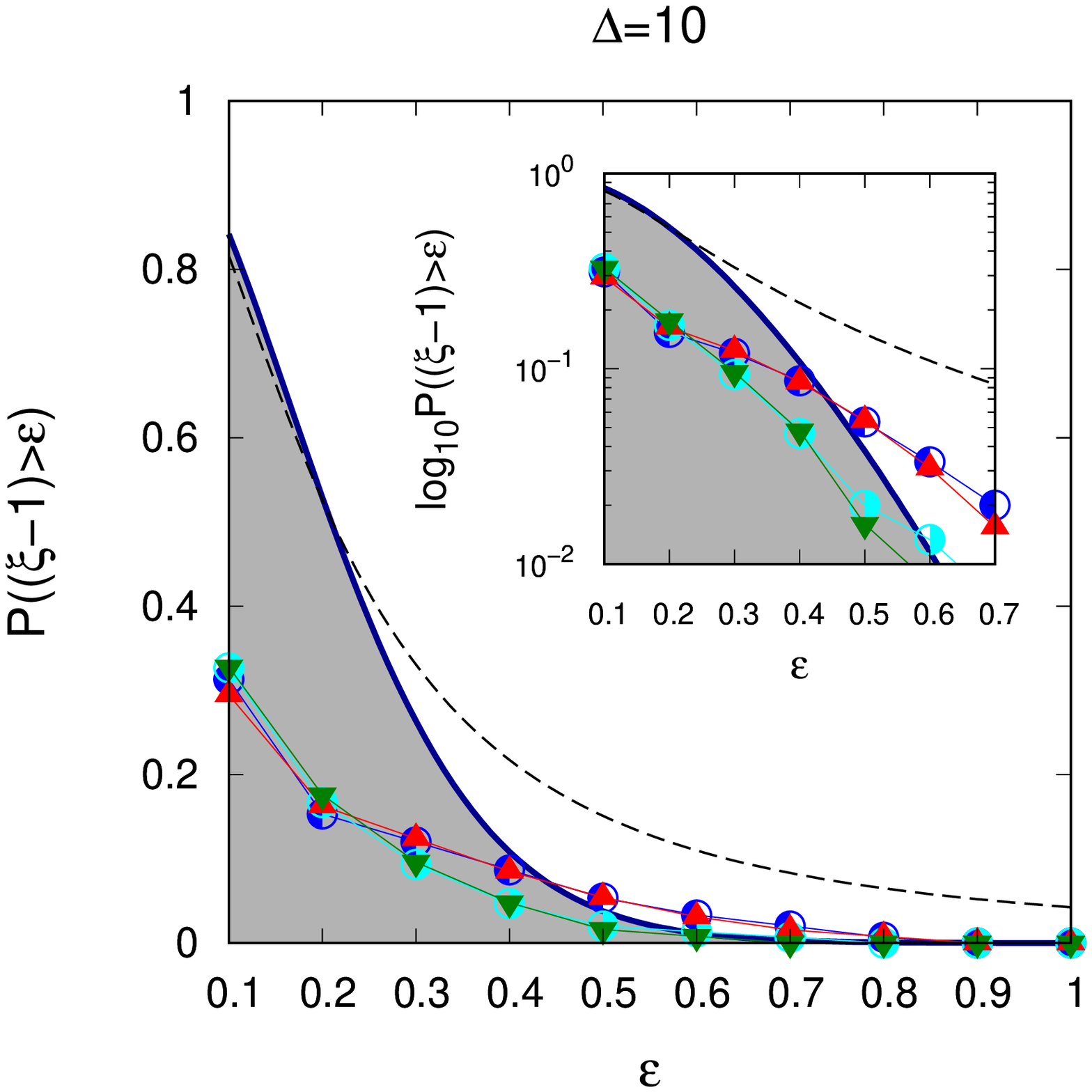}
\includegraphics[scale=0.301,angle=270]{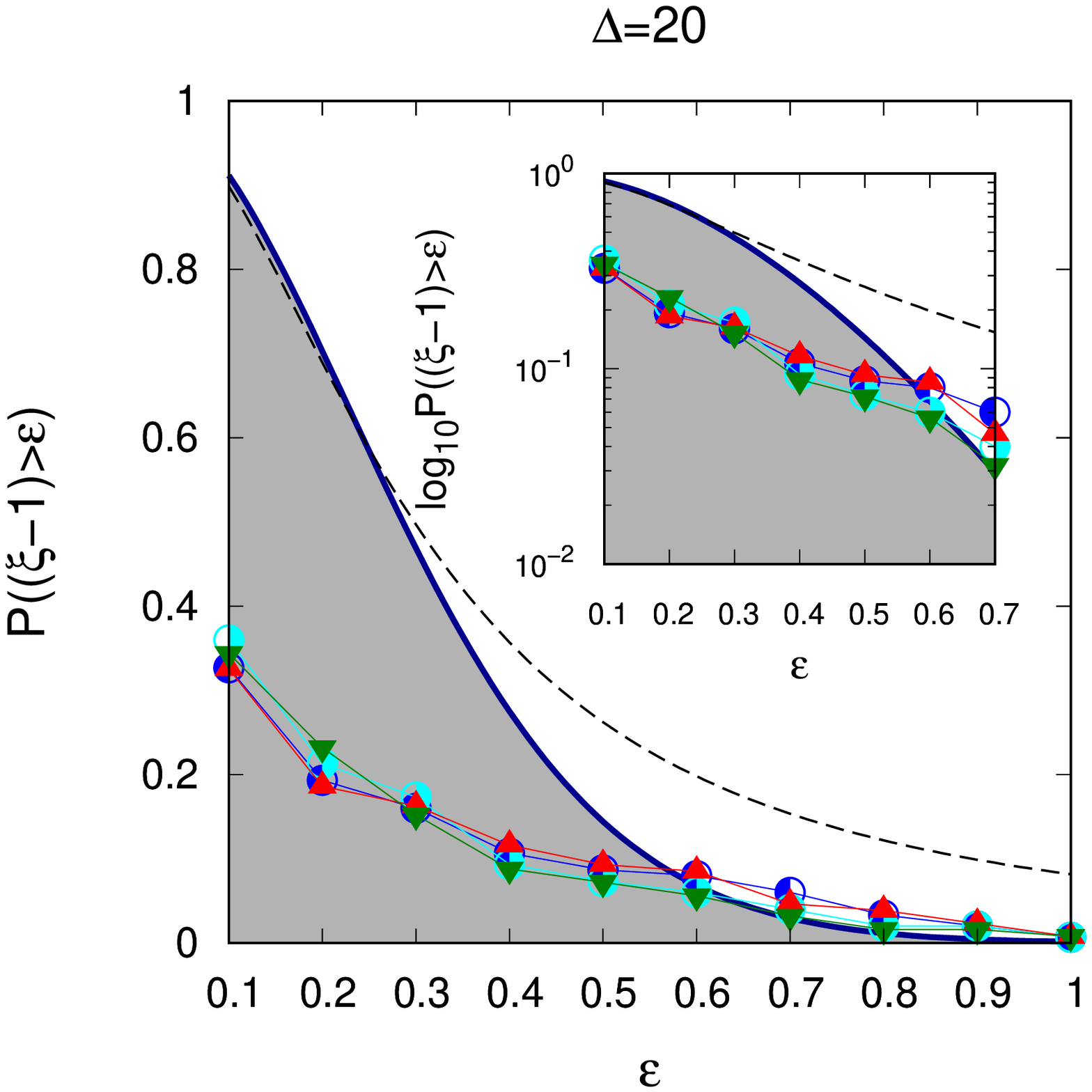}
\caption{Variation of the estimates of $P\left((\xi-1)>\epsilon\right)$ with
respect to $\epsilon$, for the datasets of polystyrene beads tracked in aqueous
solution. The figure shows the results from the set of trajectories filtered
using the JB test (labeled "BM,filtered,x-dim" and "BM,filtered,y-dim"), as
compared to the full set of trajectories (labeled "BM,x-dim" and "BM,y-dim").
The parameters $N=300$ for all the datasets while $M=150$, $129$, and $125$ for
the unfiltered sets, dataset labeled "BM,filtered,x-dim" and the dataset labeled
"BM,filtered,y-dim" respectively.}
\label{figs9}
\end{figure*}

\end{appendix}

\end{document}